\documentclass[11pt,letterpaper]{article}
\usepackage{amsmath,amssymb,amsfonts}
\usepackage{graphicx,wrapfig}
\usepackage{./files/jheppubcb}
\usepackage[usenames,dvipsnames]{xcolor}
\usepackage{skak}

\graphicspath{{./figures/}}
\numberwithin{equation}{section}
\allowdisplaybreaks

\newcommand{\be}{\begin{equation}}
\newcommand{\ee}{\end{equation}}
\newcommand{\ba}{\begin{aligned}}
\newcommand{\ea}{\end{aligned}}
\newcommand{\ben}{\begin{eqnarray}\displaystyle}
\newcommand{\een}{\end{eqnarray}}

\newcommand{\wt}{\widetilde}

\newcommand{\ie}{\emph{i.e.}}
\newcommand{\cf}{\emph{cf.}}

\newcommand{\Tr}{{\rm Tr\,}}
\newcommand{\ND}{{\rm ND}}

\renewcommand{\(}{\left(}
\renewcommand{\)}{\right)}
\renewcommand{\[}{\left[}
\renewcommand{\]}{\right]}

\renewcommand{\t}{\underline{t}}
\newcommand{\w}{w}

\newcommand{\ve}{\varepsilon}

\newcommand{\cA}{\mathcal{A}}
\newcommand{\cB}{\mathcal{B}}
\newcommand{\cC}{\mathcal{C}}
\newcommand{\cD}{\mathcal{D}}

\newcommand{\cI}{\mathcal{I}}
\newcommand{\cJ}{\mathcal{J}}

\newcommand{\cM}{\mathcal{M}}
\newcommand{\cN}{\mathcal{N}}
\newcommand{\cO}{\mathcal{O}}

\newcommand{\cQ}{\mathcal{Q}}

\newcommand{\cS}{\mathcal{S}}

\newcommand{\Z}{{\mathbb Z}}
\newcommand{\R}{{\mathbb R}}
\newcommand{\C}{{\mathbb C}}
\newcommand{\e}{\epsilon}
\renewcommand{\P}{{\mathbb P}}

\setboardfontsize{8}

\title{The superconformal index of $\cN=1$ class $\cS$ fixed points}

\author[\BlackKingOnWhite]{Christopher Beem}
\affiliation[\BlackKingOnWhite]{Simons Center for Geometry and Physics\\ 
State University of New York\\ 
Stony Brook, NY 11794-3636, USA}
\emailAdd{cbeem@scgp.stonybrook.edu}

\author[\BlackKnightOnWhite]{and Abhijit Gadde}
\affiliation[\BlackKnightOnWhite]{California Institute of Technology \\ Pasadena, CA 91125, USA}
\emailAdd{abhijit@caltech.edu}

\abstract{%
We investigate the superconformal index of four-dimensional $\cN=1$ superconformal field theories that arise on coincident M5 branes wrapping a holomorphic curve in a local Calabi-Yau three-fold. The structure of the index is very similar to that which appears in the special case preserving $\cN=2$ supersymmetry. We first compute the index for the fixed points that admit a known four-dimensional ultraviolet description and prove infrared equivalence at the level of the index for all such constructions. These results suggest a formulation of the index as a two-dimensional topological quantum field theory that generalizes the one that computes the $\cN=2$ index. The TQFT structure leads to an expression for the index of all class $\cS$ fixed points in terms of the index of the $\cN=2$ theories. Calculations of spectral data using the index suggests a connection between these families of fixed points and the mathematics of $SU(2)$  Yang-Mills theory on the wrapped curve.
}
\begin{document}
\maketitle

\section{Introduction}\label{sec:intro}

In recent years, a unifying principle has emerged in the study of many supersymmetric gauge theories. The principle holds that for theories that are realized by compactifying a six-dimensional $(2,0)$ theory on a compact manifold, many (if not all) of the protected quantities in the theories can be reformulated in terms of properties of the compactification manifold. In the case of four-dimensional theories with $\cN=2$ supersymmetry, these are the theories of class $\cS$, which are labeled by a Riemann surface known as the \emph{UV curve} \cite{Gaiotto:2009we,Gaiotto:2008cd}. In the case of three-dimensional theories with $\cN=2$ supersymmetry, a large class of \emph{three-manifold theories} have been described which arise from compactification on a cusped hyperbolic three-manifold \cite{Dimofte:2011ju} (see also \cite{Cecotti:2011iy}). It seems quite possible that theories with such a six-dimensional origin may fill out a significant portion of the space of all field theories in dimension four and lower with sufficient supersymmetry.

The theories that have received the most attention so far are really just the tip of an iceberg. It was shown in \cite{Bah:2012dg,Bah:2011vv} (following the work of \cite{Benini:2009mz}) that there exists a vast generalization of the class $\cS$ theories in four dimensions to theories that preserve only four supercharges. These are the theories realized on coincident M5 branes that wrap a complex curve in a local Calabi-Yau three-fold, and for fixed curve topology there are \emph{infinitely many} fixed points, of which the $\cN=2$ theory is the special case for which the Calabi-Yau is two complex dimensional. Indeed, the proliferation of fixed points with less supersymmetry is ubiquitous in the industry of compactifying the $(2,0)$ theory, and leads to extra fixed points in three dimensions with $\cN=1$ supersymmetry \cite{Acharya:2001aa} and in two dimensions with $\cN=(0,2)$ supersymmetry \cite{BeniniBobev,Benini:2012aa}. These less-supersymmetric systems have hardly been explored, and it remains to be seen how much geometric structure they share with their highly supersymmetric cousins.

In this note, we present a successful first venture into this less supersymmetric landscape by computing the superconformal index of the $\cN=1$ class $\cS$ fixed points of \cite{Bah:2012dg,Bah:2011vv}. The resulting structure is described in terms of a two-dimensional topological quantum field theory (TQFT) on the UV curve of the theory. By analyzing our results, we are able to determine the number of low-lying protected operators in these SCFTs, which confirms some predictions in the original papers and resolves a puzzle as well. Moreover, for a large subset of these theories (termed the \emph{accessible} theories in this paper), there exist a number of inequivalent four-dimensional UV constructions for the same IR fixed point. Understanding the equivalence of these constructions at the level of the superconformal index makes crucial use of the mathematical machinery that underlies several known $\cN=1$ dualities, suggesting that class $\cS$  may play a unifying role for $\cN=1$ dualities just as it did for $S$-dualities of $\cN=2$ theories \cite{N1dualities}.

The organization of the note is as follows: In Section \ref{sec:b3wrev}, we review the construction of $\cN=1$ fixed points from wrapped M5 branes as well as the four-dimensional construction of accessible theories. In Section \ref{sec:indexrev}, we describe the $\cN=1$ superconformal index and determine what aspects of the spectrum of an SCFT are determined unambiguously by the index. In Section \ref{sec:SU(2)index}, we compute the index for accessible fixed points at rank one, and in doing so we verify the universality of the IR limit for different UV constructions. In Section \ref{sec:TQFT}, we describe the generalization to higher rank theories in the language of a two-dimensional topological quantum field theory. The structure of the TQFT also leads to a prediction for the index of inaccessible theories. In Section \ref{sec:simplelimits}, we describe limits of the $\cN=1$ index where there is a simplification of the structure constants of the TQFT.

\section{$\cN=1$ class $\cS$ fixed points}\label{sec:b3wrev}

This section is a brief review of the four-dimensional superconformal fixed points that arise on M5 branes. For the detailed description, see \cite{Bah:2012dg}. The idea is to consider a local Calabi-Yau three-fold geometry that is a decomposable $\C^2$ bundle over a complex curve with genus $g$ and $h$ punctures, the UV curve:\footnote{While the original work on these theories focused on only unpunctured UV curves, we see in this paper that the incorporation of punctures in the index is natural. The physical nature of the $\cN=1$ punctures appearing in this work deserves further attention. }
\be\label{CYgeom}
X = L_1\oplus L_2\rightarrow\cC_{g,h}~.
\ee
The Calabi-Yau condition in the neighborhood of the zero section is that the determinant line bundle be equal to the canonical bundle $K_{\cC_{g,h}}$ of the curve. This restricts the vector bundle to be of the specialized form,
\be
X = L\oplus (K_{\cC}\otimes L^*)\rightarrow\cC_{g,h}~.
\ee 
Topologically, this specialization implies that the Chern numbers $\ell_1$ and $\ell_2$ of the two line bundles are related as follows:
\be\label{CYcond}
\ell_1+\ell_2=2g+h-2~.
\ee
The topologically distinct choices of line bundles can be parameterized by a rational \emph{twist parameter}:
\be
z=\frac {\ell_1-\ell_2}{2g+h-2}~.
\ee

When $N$ coincident M5 branes wrap the curve, their infrared dynamics are described by an $\cN=1$ SCFT that we denote $T_N[\,\cC_{g,h}\,;z\,]$. It is an important fact that the local geometry $X$ admits a $U(1)_1\times U(1)_2$ isometry, with each factor acting by rotations of the fiber of the line bundle indexed by the same number. The covariantly constant spinor on $X$, which generates the supersymmetry transformation for the partially twisted field theory on the M5-branes, has  charge one under a symmetry $R_0$ that acts on the line bundles with charges $(+\tfrac 12,+\tfrac 12)$. There is an additional flavor symmetry, denoted $F$, that acts on the line bundles with charges $(-\tfrac 12,+\tfrac 12)$ and  with respect to which the supercharges are neutral.\footnote{The canonical flavor symmetry $F$ used in this note differs from that of \cite{Bah:2012dg} by $F_{\rm BG}=-F_{{\rm B}^3{\rm W}}$.} At the fixed point, the superconformal R-symmetry is a combination of these,
\be\label{Rsc}
R_{\rm sc}=R_0+\ve F~,
\ee
where the real coefficient $\ve$ is fixed by $a$-maximization \cite{amax}.

This six-dimensional construction is not presently very useful for the purpose of computing the superconformal index of the fixed points. Fortunately, a subset of these theories can be constructed purely in four-dimensional terms. The theories for which this method applies are those with $\ell_1,\ell_2\geq0$, and we refer to these as \emph{accessible} theories.%
\footnote{This definition of accessible theories precludes the exceptional case of $\cC=\P^1$. Additionally, for $\cC=T^2$, the only accessible theory is $\cN=4$ super Yang-Mills which, though it certainly \emph{is} accessible, is also an exceptional case due to enhanced supersymmetry. In general, we focus on hyperbolic UV curves. It is made clear when we are discussing special cases.}
The remaining theories are, in this sense, \emph{inaccessible}. The procedure for generating a four-dimensional UV description of an accessible fixed point is as follows:
\begin{itemize}
\item Choose a pants decomposition of $\cC_{g,h}$ into $2g+h-2$ trinions. 
\item Deform the line bundles $L_1$ and $L_2$ so that over each trinion they take the form%
\be
T^*\cC_{0,3}\times\C~.\notag
\ee 
This amounts to dividing the trinions into two groups. For $\ell_1$ of the trinions, $L_1$ plays the role of the cotangent bundle, and we describe these trinions as being of type $(1,0)$. For the other $\ell_2$ trinions, $L_2$ is the cotangent bundle, and they are described as being of type $(0,1)$. We then have a \emph{refined pants decomposition}.
\item Each individual trinion corresponds to a copy of the $\cN=2$ SCFT $T_{N}$. For $(1,0)$ trinions, we label the corresponding SCFT by $T_N^+$, while for $(0,1)$ trinions we label them $T_N^-$.
\item Two trinions of the same (opposite) type are glued by gauging the diagonal $SU(N)$ flavor symmetry associated to the two legs that are being glued and coupling to an ${\cN}=2$ (${\cN}=1$) vector multiplet.
\end{itemize}
The gauge interactions for the $\cN=1$ vector multiplets in such a construction are asymptotically free, and the theory flows to strong coupling in the IR. The IR fixed point is the SCFT of interest. For a given choice of UV curve and twist parameter, there will generally be several inequivalent UV constructions. These are different theories, but they are conjectured to be members of the same universality class.

In terms of the generators $R,\,r$ of $[U(1)_R\subset SU(2)_R]\times U(1)_r$ for a $T_N$ SCFT or an $\cN=2$ vector multiplet, the symmetries of the Calabi-Yau act as\footnote{The normalization here is such that supercharges have half-integral charges under both $r$ and $R$.}
\be\label{TNsymmetrymap}
R_0 = R+r~,\qquad F = \pm(R-r)~.
\ee
For a constituent $T_N^{\pm}$, the plus/minus sign correlates with the sign in the action of $F$. Similarly, for an $\cN=2$ vector multiplet, the sign correlates with whether the vector gauges flavor symmetries of $T_N^+$ or $T_N^-$ constituents. For an $\cN=1$ vector multiplet, $R_0$ acts as the canonical $R$-symmetry, and $F$ does not act at all.

By construction, this recipe is only well-defined for accessible theories. No four dimensional UV description has been suggested for the inaccessible fixed points. Nevertheless, we will see that the superconformal index for the accessible theories suggests a natural extension that encompasses \emph{all} fixed points.

For later reference, we remind the reader of some of the properties of these fixed points that were reported in the original papers on the subject. Implementing $a$-maximization fixes $\ve$ in Equation \eqref{Rsc}. The result is that $\ve$ vanishes for $z=0$ and varies monotonically as a function of $z$ for fixed $N$, becoming positive for $z>0$ and negative for $z<0$ For all $z$ and $N$, we find
\be\label{epsilonbound}
|\ve|<\frac{1}{\sqrt3}~.
\ee
Furthermore, For fixed topology of the UV curve and fixed $N$, the central charges $a$ and $c$ both increase monotonically in $|z|$, with linear asymptotics. 

For the exceptional case $\cC=\P^1$, there is no fixed point for $z=0,\pm1$ with only the UV symmetries. Indeed, the dynamics on these brane configurations is expected to be confining in the infrared \cite{Maldacena:2000mw}. Also in this case it was observed in \cite{Bah:2012dg} that for $|z|=N=2$, the resulting central charges violate the bounds of \cite{Hofman:2008ar,Kulaxizi:2010jt} for the ratio $a/c$. When $\cC=T^2$, the fixed point at $z=0$ is $\cN=4$ SYM, which has enhanced supersymmetry and is not well-described by these methods.

For all hyperbolic UV curves, the choice $z=0$ leads to a special fixed point which was studied already in \cite{Benini:2009mz}. This is the same fixed point that is attained by starting with the $\cN=2$ theory for the same UV curve and deforming by holomorphic mass terms for the adjoint-valued chiral multiplets in $\cN=2$ vector multiplets. For these theories, the symmetry $F$ is enhanced to $SU(2)_F$ on a sublocus of the conformal manifold, but is broken elsewhere.

The conformal manifold of the theories has been analyzed holographically and for the accessible fixed points, where it was found that the number of exactly marginal operators is given by
\be\label{conformalmanifold}
\dim_\C \cM_{\rm CFT}=\begin{cases} 
\mbox{$6g-6$}&	\qquad 	g>1,  \;z=0~, 		\\
\mbox{$4g-3$}&	\qquad 	g>1,  \;z\neq0~, 	\\
\mbox{$2$} 	&	\qquad 	g = 1,\;z\neq0~,		\\
\mbox{$0$} 	&	\qquad 	g = 0,\;z>1~.			
\end{cases}
\ee

On the basis of the six-dimensional construction of the theories, the results for $z\neq0$ are expected to remain the same even for the inaccessible fixed points. The authors of \cite{Bah:2012dg} also left an unresolved puzzle regarding the number of relevant operators: A na\"ive counting of relevant chiral operators seemed to show a mismatch between different UV constructions. This puzzle is resolved in terms of the index of these theories later in this note.

\section{The $\cN=1$ superconformal index}\label{sec:indexrev}

The superconformal index is a powerful invariant of superconformal field theories that counts protected states in radial quantization. For $\cN=1$ theories in four dimensions there are both left-handed and right-handed indices that count states that are annihilated by the supercharges $\cQ=\cQ_{-}$ and $\cQ=\wt\cQ_{\dot{-}}$, respectively. These indices can be written as trace formul\ae\ and there are two principle sets of conventions for the fugacities
\be\ba\label{indexdef}
\cI^{\rm L} &= \Tr(-1)^Ft^{2(E+\cJ_1)}y^{2\cJ_2}\prod a_i^{e_i}=\Tr(-1)^Fp^{\cJ_{12}-\frac 12 r}q^{\cJ_{34}-\frac 12 r}\prod a_i^{e_i}~,\\
\cI^{\rm R} &= \Tr(-1)^Ft^{2(E+\cJ_2)}y^{2\cJ_1}\prod a_i^{e_i}=\Tr(-1)^Fp^{\cJ_{12}+\frac 12 r}q^{-\cJ_{34}+\frac 12 r}\prod a_i^{e_i}~,
\ea\ee
where $E$ is the conformal Hamiltonian, $\cJ_{12}=\cJ_2+\cJ_1$ and $\cJ_{34}=\cJ_2-\cJ_1$ in terms of the Cartan generators of $SU(2)_1$ and $SU(2)_2$, and $e_i$ are the generators of a maximal torus of the global flavor symmetry group. The two sets of fugacities are related by $p=t^3y$, $q=t^3y^{-1}$, and the two expressions are equivalent because the indices receive non-vanishing contributions only from states that satisfy an appropriate chirality condition,
\be\label{chirality}
\{\,\cQ\,,\,\cQ^\dagger\,\}\,|\psi\rangle = 0\quad\implies\quad \Delta-2j_{1,2}\pm\frac 32 r = 0~.
\ee
Otherwise, pairs of states $|\psi\rangle$ and $\cQ\,|\psi\rangle$ cancel out of the index. 

The states of an $\cN=1$ SCFT are organized into representations of the superconformal algebra $SU(2,2\,|\,1)$, and only certain shortened representations of this algebra make non-vanishing contributions to the index. We now review these short representations and their corresponding contributions to the superconformal index. For a more extensive discussion, see \cite{Dolan:2002zh,Kinney:2005ej} and Appendix A of \cite{Gadde:2010en}.

\subsection{Superconformal representation theory and the index}

A generic long multiplet $\cA_{r(j_{1},j_{2})}^{\Delta}$ of ${\cN}=1$ superconformal algebra is generated by the action of the four Poincar\'e supercharges $(\cQ_\alpha,\wt\cQ_{\dot{\alpha}})$ on a superconformal primary state, which by definition is annihilated by superconformal charges $(\cS_{\alpha},\wt\cS_{\dot\alpha})$. The multiplet is characterized by the charges $(\Delta,r,j_1,j_2)$ of the primary with respect to the charges $(E,r,\cJ_1,\cJ_2)$. The absence of negative norm states in the multiplet imposes certain inequalities on these quantum numbers,
\be\begin{split}\label{unitarity}
\Delta~&\geq~ 2-2\delta_{j_1,0}+2j_1-\tfrac 32 r~,\\
\Delta~&\geq~ 2-2\delta_{j_2,0}+2j_2+\tfrac 32 r~, \\
\Delta~&\notin \(-\tfrac 32 r\,,\,2-\tfrac 32 r\)~,\qquad\qquad\qquad\!\!\!\text{if}~~j_1=0~,\\
\Delta~&\notin \(\tfrac 32 r\,,\,2+\tfrac 32 r\)~,~\qquad\qquad\qquad\text{if}~~j_1=0~,\\
\Delta~&\geq~ 2+j_1+j_2~,~~~\qquad\qquad\qquad\text{if}~~j_1\neq0,~~j_2\neq0~,\\
\Delta~&\geq~ 1+j_1+j_2~,~~~\qquad\qquad\qquad\text{if}~~j_1=0~~\text{or}~~j_2=0~.
\end{split}\ee
When these inequalities are saturated, some combination of the Poincar\'e supercharges will annihilate the primary as well, resulting in a shortened multiplet. The relevant property of these short multiplets is that they must always saturate the unitarity bound in order to be free of negative normed states, and so their conformal dimension is fixed in terms of other quantum numbers and is protected against corrections as one changes the parameters of the theory.

The possible shortening conditions of the ${\cN}=1$ superconformal algebra are summarized in Table \ref{N1-shortening}. The $\cD$ and $\bar{\cD}$ multiplets correspond to free fields and do not play a role in the theories analyzed in this paper.
\begin{center}
\begin{table}[h!]
{\small
\begin{centering}
\begin{tabular}{|l|l|l|l|l|}
\hline
\multicolumn{4}{|c|}{Shortening Conditions} & Multiplet\\
\hline
$\cB$ & $\cQ_{\alpha}|r\rangle^{h.w.}=0$ & $j_1=0$ & $\Delta=-\frac{3}{2}r$ & $\cB_{r(0,j_2)}$\tabularnewline
\hline
$\bar{\cB}$ & $\bar{\cQ}_{\dot{\alpha}}|r\rangle^{h.w.}=0$ & $j_2=0$ & $\Delta=\frac{3}{2}r$ & $\bar{\cB}_{r(j_1,0)}$\tabularnewline
\hline
$\hat{\cB}$ & $\cB\cap\bar{\cB}$ & $j_1,j_2,r=0$ & $\Delta=0$ & $\hat{\cB}$\tabularnewline
\hline
$\cC$ & $\e^{\alpha\beta}\cQ_{\beta}|r\rangle_{\alpha}^{h.w.}=0$ &  & $\Delta=2+2j_1-\frac{3}{2}r$ & $\cC_{r(j_1,j_2)}$\tabularnewline
 & $(\cQ)^{2}|r\rangle^{h.w.}=0$ for $j_1=0$ &  & $\Delta=2-\frac{3}{2}r$ & $\cC_{r(0,j_2)}$\tabularnewline
\hline
$\bar{\cC}$ & $\e^{\dot{\alpha}\dot{\beta}}\bar{\cQ}_{\dot{\beta}}|r\rangle_{\dot{\alpha}}^{h.w.}=0$ &  & $\Delta=2+2j_2+\frac{3}{2}r$ & $\bar{\cC}_{r(j_1,j_2)}$\tabularnewline
 & $(\bar{\cQ})^{2}|r\rangle^{h.w.}=0$ for $j_2=0$ &  & $\Delta=2+\frac{3}{2}r$ & $\bar{\cC}_{r(j_1,0)}$\tabularnewline
\hline
$\hat{\cC}$ & $\cC\cap\bar{\cC}$ & $\frac{3}{2}r=(j_1-j_2)$ & $\Delta=2+j_1+j_2$ & $\hat{\cC}_{(j_1,j_2)}$\tabularnewline
\hline
$\cD$ & $\cB\cap\bar{\cC}$ & $j_1=0,-\frac{3}{2}r=j_2+1$ & $\Delta=-\frac{3}{2}r=1+j_2$ & $\cD_{(0,j_2)}$\tabularnewline
\hline
$\bar{\cD}$ & $\bar{\cB}\cap\cC$ & $j_2=0,\frac{3}{2}r=j_1+1$ & $\Delta=\frac{3}{2}r=1+j_1$ & $\bar{\cD}_{(j_1,0)}$\tabularnewline
\hline
\end{tabular}
\par\end{centering}
} \caption{\label{N1-shortening}Shortening conditions for the $SU(2,2\,|\,1)$ superconformal algebra.}
\end{table}
\par\end{center}\vspace{-15pt}
If the charges of a collection of short multiplets obey certain relations, they can combine to form a long multiplet which is no longer protected. Alternatively, one can understand this recombination in reverse, as a long multiplet decomposing into a collection short multiplets as the conformal dimension of its primary hits the BPS bound. This phenomenon plays a crucial role in extracting spectral information about an SCFT from its index because the index counts short multiplets of the theory \emph{up to recombination}. The collective contributions to the index from short multiplets that can recombine vanishes. The recombination equations for ${\cN}=1$ superconformal algebra are as follows:
\ben\label{recombination}
\cA_{r(j_{1},j_{2})}^{2+2j_{1}-\frac{3}{2}r} & \longrightarrow & \cC_{r(j_{1},j_{2})}\oplus\cC_{r-1(j_{1}-\frac{1}{2},j_{2})}~,\notag\\
\cA_{r(j_{1},j_{2})}^{2+2j_{2}+\frac{3}{2}r} & \longrightarrow & \bar{\cC}_{r(j_{1},j_{2})}\oplus\bar{\cC}_{r+1(j_{1},j_{2}-\frac{1}{2})}~,\\
\cA_{\frac{2}{3}(j_{1}-j_{2})(j_{1},j_{2})}^{2+j_{1}+j_{2}} & \longrightarrow & \hat{\cC}_{(j_{1},j_{2})}\oplus\cC_{\frac{2}{3}(j_{1}-j_{2})-1(j_{1}-\frac{1}{2},j_{2})}\oplus\bar{\cC}_{\frac{2}{3}(j_{1}-j_{2})+1(j_{1},j_{2}-\frac{1}{2})}~.\notag
\een
For our purposes, $\cB$ multiplets can be treated formally as a special case of $\cC$ multiplets with unphysical spin quantum numbers,
\be
\cB_{r(0,j_2)}=:\cC_{r+1(-\frac 12,j_2)}~,\qquad\qquad\bar{\cB}_{r(j_{1},0)}=:\bar{\cC}_{r-1(j_1,-\frac 12)}~.
\ee
Thus our discussions can be phrased entirely in terms of ${\cC}$ type multiplets. 

\vskip15pt
\subsection*{An example of recombination}
One example of recombination that will concern us is for the long multiplet $\cA_{0(0,0)}^{2+\epsilon}$ as $\epsilon\to0$. The multiplet hits the BPS bound and splits into three short multiplets according to the third rule in \eqref{recombination}:
\be
\cA_{0(0,0)}^{2}\longrightarrow\hat{\cC}_{(0,0)}\oplus\cC_{-1(-\frac{1}{2},0)}\oplus\bar{{\cC}}_{1(0,-\frac{1}{2})}=\hat{{\cC}}_{(0,0)}\oplus({\cB}_{-2(0,0)}\oplus \bar{\cB}_{2(0,0)})
\ee
The multiplet $\hat{\cC}_{(0,0)}$ contains a conserved current, while the multiplet $\cB_{-2(0,0)}$ contains a chiral primary $\cO$ of dimension three and an associated marginal F-term deformation $\int d^{2}\theta\,\cO$. The recombination described above demonstrates the fact that a marginal operator can fail to be exactly marginal if and only if it combines with a conserved current corresponding to a broken global symmetry. This particular recombination and its implications for the space of exactly marginal deformations of an SCFT has been studied in detail in \cite{Green:2010da}.
\vspace{15pt}

The $\cC$ ($\bar{\cC}$) multiplets contribute only to the left-handed index (right-handed index), while $\hat\cC$ multiplets contribute to both. We restrict our attention to $\cI^{{\tt L}}$ and treat $\hat \cC$ as a special case of $\cC$ with $r=\frac{2}{3}(j_{1}-j_{2})$. The recombination rules allow us to define equivalence classes of short representations which make identical contributions to the index,
\be\ba
\[\tilde r,j_2\]_+&:=\{\cC_{r(j_1,j_2)}\;|\;2j_1-r=\tilde r,\;\; j_1\in\Z_{\geq0}\}~,\\
\[\tilde r,j_2\]_-&:=\{\cC_{r(j_1,j_2)}\;|\;2j_1-r=\tilde r,\;\; j_1\in-\tfrac 12+\Z_{\geq0}\}~.
\ea\ee
For a $\cB$ type multiplet, the unitarity bounds of Equation \eqref{unitarity} imply that $\tilde r \geq -\frac 43+\frac 23 j_2$, while for a $\cC$ multiplet they imply $\tilde r \geq \frac 43 j_1 +\frac 23 j_2$. Consequently, there are a finite number of representations in a fixed equivalence class --- for fixed $\tilde r$, there is an upper limit on $j_1$ such that these bounds can be satisfied.

The most physically transparent set of fugacities for extracting the spectrum of an SCFT from the index is $(t,y)$, in terms of which the contribution to the left-handed superconformal index from any short multiplet in a given class is given by
\be\label{multipletindex}
\cI_{[\tilde{r},j_{2}]_{+}}^{{\tt L}}=-\cI_{[\tilde{r},j_{2}]_{-}}^{{\tt L}}=(-1)^{2j_{2}+1}\frac{t^{3(\tilde{r}+2)}\chi_{j_{2}}(y)}{(1-t^{3}y)(1-t^{3}y^{-1})}~.
\ee
We define the \emph{net degeneracy} for a given choice of $(\tilde r, j_2)$,
\be
\ND[\tilde{r},j_{2}]:=\#\;[\tilde{r},j_{2}]_{+}-\#\;[\tilde{r},j_{2}]_{-}~,
\ee
and the content of the superconformal index is encapsulated in precisely the integers $\ND[\tilde r,j_2]$. If the index of an $\cN=1$ SCFT is known, the net degeneracies can be systematically extracted by means of a \emph{sieve algorithm}:
\begin{enumerate}
\item Instantiate a set of net degeneracies $S=\{\emptyset\}$.
\item Subtract one from the index to account for the contribution of the vacuum state.
\item\label{step3} Consider the terms in the $(t,y)$ expansion with the lowest power of $t$, and decompose these into characters of $SU(2)_2$,
\be\label{leadingterm}
\cI(t,y)= (-1)^{2j_{2}+1}t^{3(\tilde{r}+2)}\(\sum_{n\in\frac12 \Z_{\geq0}} a_n\chi_{j_2=n}(y)\)+\ldots
\ee
The net degeneracy of the class $[\tilde{r},j_2=n]$ is the integer $a_n$.
\item Incorporate this information into the set of degeneracies, 
\be
S\to S\cup\{\ND[\tilde{r},j_2=n]=a_n\}~,
\ee
and update the index 
\be
\cI(t,y)\to\cI(t,y)-\frac{(-1)^{2j_{2}+1}t^{3(\tilde{r}+2)}\(\sum_{n}a_n\chi_{j_2=n}(y)\)}{(1-t^{3}y)(t^{3}y^{-1})}~.
\ee
\item Go to Step \ref{step3} and continue.
\end{enumerate}
The algorithm doesn't necessarily terminate as there are generally infinitely many short multiplets in the spectrum, but at any point in this process, the set $S$ contains the net degeneracies of all operators up to some fixed value of $\tilde r$.

\subsection{Precision operator counting}\label{subsec:opcounting}

Ideally, one would like to obtain precise information about the spectrum of an SCFT from the superconformal index in the form of actual degeneracies of certain types of operators. The most precise information comes from the equivalence classes with a small number of representatives. We now point out the two best cases available for extracting such detailed spectral data.

The optimal case is the chiral primary operators that lie in multiplets $\cB_{r(0,j_2)}$ and have $-2-\tfrac23 j_2<r\leq-\tfrac 23-\tfrac 23 j_2$. These have $\tilde r \in [-\tfrac 43+\tfrac 23j_2,\tfrac 23j_2)$, and they are the \emph{only representatives} of the equivalence class $[\tilde r,0]_-$ for this range of $\tilde r$. Furthermore, there are no unitary representations in the corresponding class $[\tilde r,0]_+$. Consequently, we can read off the exact number of such operators from the superconformal index. Specializing to $j_2=0$, these are precisely the relevant deformations of the SCFT. The number of such deformations is simply the coefficient of $t^{-3r}y^{0}$ in the index after subtracting out any non-trivial $SU(2)_2$ characters at the same power of $t$.

The next best case is for $\tilde r\in[\tfrac 23j_2,\tfrac 23+\tfrac 23j_2)$. Both $[\tilde r,j_2]_+$ and $[\tilde r,j_2]_-$ have only a single representative in this range, and so the index computes the difference in the number of such operators. For $j_2=\tilde r=0$ in particular, the representatives are $\hat{\cC}_{(0,0)}$ and $\cB_{-2(0,0)}$, respectively. The cancellation between these multiplets corresponds to precisely the recombination described in the example above, and we see that the index computes
\be\begin{split}
\ND[0,0]  &= \#\ \cB_{-2(0,0)}-\#\ \hat{\cC}_{(0,0)}\cr
 & =  \#\:\mbox{marginal operators}-\mbox{\# }\mbox{conserved currents}~.
\end{split}\ee
If all global flavor symmetries are broken at a generic point on the conformal manifold, then this net degeneracy will precisely capture the actual dimension of that conformal manifold. However, not all recombinations of the type discussed in the example necessarily take place, and in this case one must account for conserved currents in extracting the dimension of the conformal manifold. Again, this net degeneracy is easily computed by expanding the index to order $t^6$ and subtracting out all nontrivial characters for $SU(2)_2$.

For $\tilde r\geq\tfrac 23$, there will be several representatives that are indistinguishable to the index, and the cancellations among them do not correspond to any obvious physical phenomenon such as symmetry breaking. Thus, the most immediate spectroscopic use of the index is the analysis of relevant and marginal operators at a fixed point.

\section{The index of accessible rank one theories}\label{sec:SU(2)index}

There is a subset of the fixed points described in Section \ref{sec:b3wrev} that admits Lagrangian UV descriptions: the accessible fixed points for two M5 branes. In this case, the $T_N$ building block used in the four-dimensional UV construction is simply the theory of eight free chiral multiplets transforming in the trifundamental representation of an $SU(2)^3$ global symmetry group. The theories in question are constructed from a finite number of such trifundamentals, which are coupled by $\cN=2$ and $\cN=1$ vector multiplets that gauge subgroups of the global symmetry. Crucially, this construction preserves the the full $U(1)_1\times U(1)_2$ global symmetry of the fixed points, and the index can be computed by using the prescription of \cite{Romelsberger:2005eg, Romelsberger:2007ec,Dolan:2008qi}. Below we describe the computation in general, and then the consider several examples to illustrate. We will describe the computations in the section using $(p,q)$ fugacities, which are the ones for which these indices have a natural expression in terms of elliptic gamma functions.

\subsection{Ultraviolet computation of the index}

Invariance of the index under marginal deformations facilitates direct computation of the index for many of the classic $\cN=2$ SCFTs that possess a free limit point on their conformal manifold. However, $\cN=1$ fixed points tend to be irreparably interacting. Nevertheless, there exists a prescription for computing the index of such SCFTs when there is a known weakly coupled UV description of the theory that flows to the fixed point in question. This is because, for the purposes of counting states obeying the chirality condition \eqref{chirality}, one does not need superconformal symmetry, but rather the only requirement is the existence of supercharges $\cQ$ and $\cQ^\dagger$ that generate a symmetry of the theory on $S^3\times\R$. Away from conformality, there is no canonical choice of $R$-symmetry, and the index depends on the choice.

At a conformal fixed point, the states organize themselves into representations of the four-dimensional $\cN=1$ superconformal algebra $SU(2,2\,|\,1)$, but if the index is computed in the UV, the correct superconformal $R$-symmetry must be identified in order to extract information about superconformal representations. It is therefore crucial that there be no accidental symmetries in the IR that mix with the superconformal $R$-symmetry. The prescription of R\"omelsberger is to compute the supersymmetric index at a UV fixed point (possibly free), and to substitute the appropriate R-charges to find the index of the IR fixed point. In the case at hand, this means that we will be using the indices of free $\cN=2$ and $\cN=1$ vector multiplets and free trifundamentals. The index that we want to compute is defined as
\be
\cI = \Tr(-1)^Fp^{\cJ_{12}-\frac 12 R_{\rm sc}}q^{\cJ_{34}-\frac 12 R_{\rm sc}}x^F=\Tr(-1)^Fp^{\cJ_{12}-\frac 12 R_0}q^{\cJ_{34}-\frac 12 R_0}(x(pq)^{-\frac \ve2})^F
\ee
so it will be sufficient to compute the simpler index,
\be\label{simpleindex}
\Tr(-1)^Fp^{\cJ_{12}-\frac 12 R_0}q^{\cJ_{34}-\frac 12 R_0}\xi^F~,
\ee
from which the true superconformal index can be recovered by redefining $\xi=x(pq)^{-\frac\ve2}$ for the value of $\ve$ that is determined by $a$-maximization.

This index is known for all of the relevant building blocks of the UV theories in question, \ie, trifundamentals of types ``$+$'' and ``$-$'', $\cN=2$ vectors of type ``$+$'' and ``$-$'', and $\cN=1$ vectors. The symmetry $F$ in \eqref{simpleindex} acts differently on the building blocks of different sign, which is where the choice of line bundles in \eqref{CYgeom} enters into the index computation. The building block are all free fields, and the index of a collection of free fields $\{\phi\}$ can be expressed simply in terms of ``single letter indices'' by means of the plethystic exponential,
\be
\cI_{\{\phi\}}(t,y,\ldots)={\rm PE}\[i_{\{\phi\}}(\vec a,\xi;p,q)\] := \exp\(\sum_{n=1}^\infty\frac{1}{n}i_{\{\phi\}}(\vec a^{\,n},\xi^n;p^n,q^n)\)~.
\ee
The relevant single-letter indices are listed below:%
\footnote{\label{n2indexfoot}For $\cN=2$ building blocks, these $\cN=1$ indices are related in a simple way to the more canonical $\cN=2$ index, which is defined as
\be
\cI_{\cN=2}(p,q,\t)=\Tr(-1)^Fp^{\cJ_{12}-\frac 12r_{\cN=2}}q^{\cJ_{34}-\frac 12r_{\cN=2}}\t^{R-r_{\cN=2}}~,
\ee
where $r_{\cN=2}$ is the generator of $U(1)_r$ and $R$ is the Cartan generator of $SU(2)_R$. The relation is a simple modification at the level of the fugacities,
\be
\cI_{\cN=1}^{+}(\vec a,\xi;p,q)=\cI_{\cN=2}(\vec a;p,q,(pq)^{\frac 12}\xi)~,\qquad \cI_{\cN=1}^{-}(\vec a,\xi;p,q)=\cI_{\cN=2}(\vec a;p,q,(pq)^{\frac 12}\xi^{-1})~.
\ee
}
\ben\label{singleletters}
i_{V}^{+}(a,\xi;p,q) & =&  \frac{-\sqrt{pq}(\xi-\xi^{-1})-p-q+2pq}{(1-p)(1-q)}\chi_{adj}(a)~,\\
i_{V}^{-}(a,\xi;p,q) & =&  \frac{\sqrt{pq}(\xi-\xi^{-1})-p-q+2pq}{(1-p)(1-q)}\chi_{adj}(a)~,\\
i_{V}^{\emptyset}(a,\xi;p,q) & =&  \frac{-p-q+2pq}{(1-p)(1-q)}\chi_{adj}(a)~,\\
i_{T_2}^{+}(a_1,a_2,a_3,\xi;p,q) & =&  \frac{(pq)^{\frac 14}\xi^{\frac 12}-(pq)^{\frac 34}\xi^{-\frac 12}}{(1-p)(1-q)}\chi_{\square}(a_1)\chi_{\square}(a_2)\chi_{\square}(a_3)~,\\
i_{T_2}^{-}(a_1,a_2,a_3,\xi;p,q) & =&  \frac{(pq)^{\frac 14}\xi^{-\frac 12}-(pq)^{\frac 34}\xi^{\frac 12}}{(1-p)(1-q)}\chi_{\square}(a_1)\chi_{\square}(a_2)\chi_{\square}(a_3)~.
\een
Conveniently, the plethystic exponent of these single letter indices admit a nice expression in terms of elliptic gamma functions,
\be\begin{split}\label{SU(2)indices}
\cI_V^{+}(a,\xi;p,q) & =  \frac{(p;p)(q;q)}{(1-a^2)(1-a^{-2})}\frac{1}{\Gamma(a^{\pm2};p,q)}\frac{1}{\Gamma((pq)^{\frac 12}\xi a^{\pm2};p,q)}~,\\
\cI_V^{-}(a,\xi;p,q) & =  \frac{(p;p)(q;q)}{(1-a^2)(1-a^{-2})}\frac{1}{\Gamma(a^{\pm2};p,q)}\Gamma((pq)^{\frac 12}\xi a^{\pm2};p,q)~,\\
\cI_V^{\emptyset}(a,\xi;p,q) & =  \frac{(p;p)(q;q)}{(1-a^2)(1-a^{-2})}\frac{1}{\Gamma(a^{\pm2};p,q)}~,\\
\cI_{T_2}^{+}(a_1,a_2,a_3,\xi;p,q) & =  \Gamma((pq)^{\frac 14}\xi^{\frac 12}a_1^{\pm}a_2^{\pm},a_3^{\pm};p,q)~,\\
\cI_{T_2}^{-}(a_1,a_2,a_3,\xi;p,q) & =  \Gamma((pq)^{\frac 14}\xi^{-\frac 12}a_1^{\pm}a_2^{\pm},a_3^{\pm};p,q)~,
\end{split}\ee
where the elliptic gamma function is defined as
\be\label{gammadef}
\Gamma(z;p,q) := \prod_{i,j=0}^{\infty} \frac{1-p^{i+1} q^{j+1}z^{-1}}{1-z p^i q^j}={\rm PE}\[\frac{z-pqz^{-1}}{(1-p)(1-q)}\]~,
\ee
and we are using a condensed notation in which $\Gamma(a^\pm;p,q):=\Gamma(a^+;p,q)\Gamma(a^-;p,q)$ and $\Gamma(a,b;p,q) := \Gamma(a;p,q)\Gamma(b;p,q)$. We will usually suppress the arguments $p,q$.

With the indices of our constituent theories in hand, it is straightforward to define the finite-dimensional contour integral that evaluates the index for any accessible $SU(2)$ theory. We now describe a few examples at relatively low genus where the integral can be arranged and evaluated to arbitrary orders in $t$, allowing us to verify infrared equivalence of different generalized quiver constructions. In addition, we read off the spectrum of relevant and marginal operators, which is the most robust physical information contained in the index (\cf\ Section \ref{subsec:opcounting}). 

\subsection{Examples for small genus}\label{subsec:examples}

Before moving on to more formal considerations, we apply the procedure described above to some simple examples. The results in this section partially resolve a puzzle from \cite{Bah:2012dg} by 
\begin{wrapfigure}{r}{7.5cm}
\centering
\includegraphics[width=6.6cm]{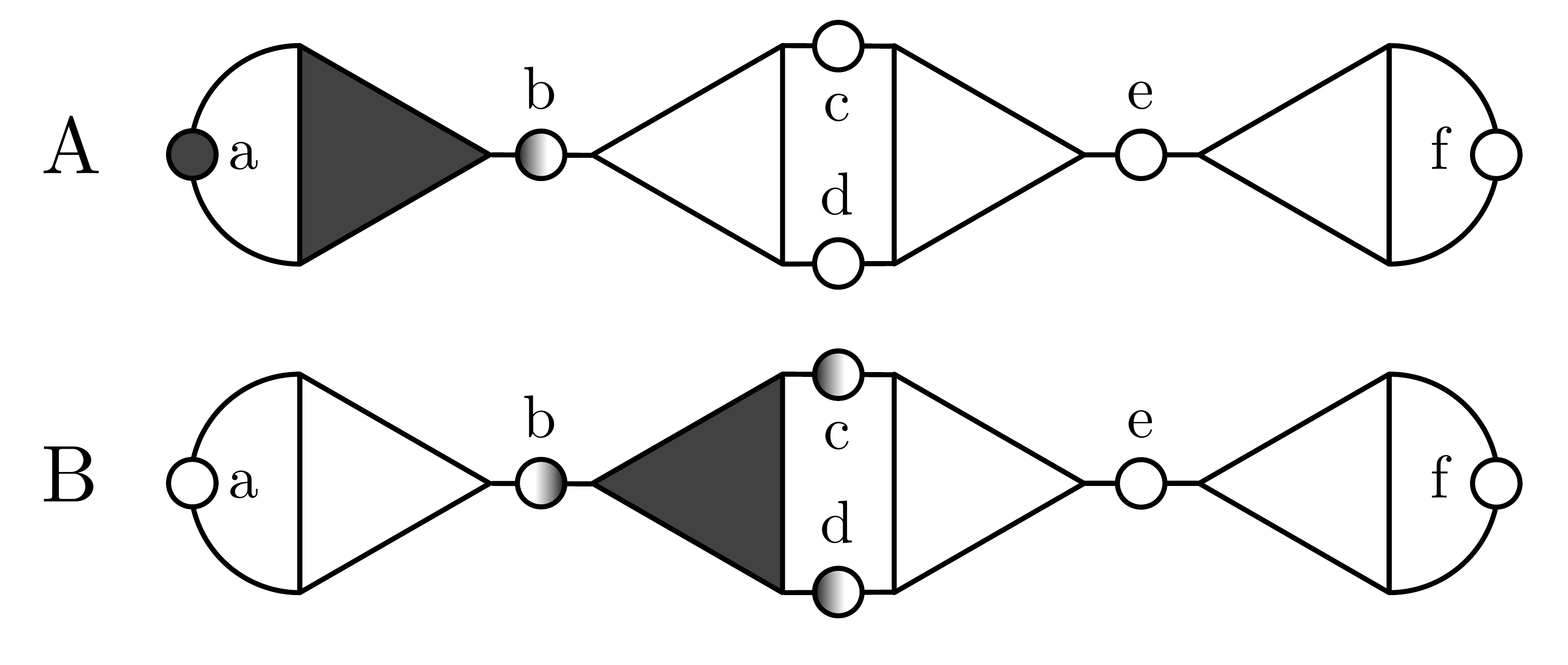}
\caption{Two generalized quivers that flow to the same SCFT. Unshaded versus shaded triangles corresponds to $T_N^+$ constituents versus $T_N^-$.}
\label{genus3}
\end{wrapfigure}
demonstrating that at rank one, the number of relevant operators matches between differentUV 
constructions. We consider theories for closed UV curves of genus three and four at accessible values of the twist parameter. For genus three, we consider $z=0$ and $\tfrac 12$, while for genus four we have $z=0,\,\tfrac 13$, and $\tfrac 23$. The setup for genus three at $z=\tfrac 12$ is described in detail, while for the remaining examples we list results and comment on their interpretation.
\subsubsection*{Genus three}
The theory at $z=\frac 12$ corresponds to the local Calabi-Yau geometry with $\ell_1=3$ and $\ell_2=1$. The four dimensional UV construction utilizes three copies of $T_N^+$ one of $T_N^-$. For illustrative purposes, we introduce two different UV constructions of the theory, which are displayed in Figure \ref{genus3} and denoted theories $A$ and $B$. Both of these constructions are expected to flow to the same fixed manifold in the infrared.

The superconformal indices of these theories can be expressed using the building blocks defined above. The result in each case is a six-dimensional contour integral,
\be\ba\label{genus3equality}
\cI_A(\xi; t,y)=&\\
\oint[da\ldots df&]{\cI}_{V}^{-}(a){\cI}_{V}^{\emptyset}(b){\cI}_{V}^{+}(c){\cI}_{V}^{+}(d){\cI}_{V}^{+}(e){\cI}_{V}^{+}(f){\cI}_{T_2}^{-}(a,a,b){\cI}_{T_2}^{+}(b,c,d){\cI}_{T_2}^{+}(c,d,e){\cI}_{T_2}^{+}(e,f,f)\notag\\
\cI_B(\xi; t,y)=&\\
\oint[da\ldots df&]{\cI}_{V}^{+}(a){\cI}_{V}^{\emptyset}(b){\cI}_{V}^{\emptyset}(c){\cI}_{V}^{\emptyset}(d){\cI}_{V}^{+}(e){\cI}_{V}^{+}(f){\cI}_{T_2}^{+}(a,a,b){\cI}_{T_2}^{-}(b,c,d){\cI}_{T_2}^{+}(c,d,e){\cI}_{T_2}^{+}(e,f,f)
\ea\ee
Here, $a,\ldots,f$ are $SU(2)$ group elements and $\oint[da]$ denotes an integral over a maximal torus of $SU(2)_{a}$ with the Haar measure. We have suppressed the arguments $\xi$ and $p,q$ that are common to all of the indices. 

Universality of the UV constructions requires $\cI_{A}=\cI_{B}$. This amounts to a certain identity for elliptic hypergeometric integrals which will be discussed in Section \ref{identities}. For now, we expand in powers of $t$ using $\mathtt{Mathematica}$ in order to count light operators and check that they agree. The first nontrivial contribution appears at $\cO(t^6)$,
\be
\cI_{A}=\cI_{B}=\cI[\cC_{3,0};z=\tfrac 12]=\(8+\xi^{-1}+4\xi^2-\(5\xi+3\xi^{-1}\)\chi_{\square}(y)\)t^6+\cO(t^9)~.
\ee
For this theory, $\ve \approx .2211$, so (after factoring out full characters of $SU(2)$ in terms of $y$) the marginal operators are counted by the coefficient of $t^6\xi^0$, and relevant operators are the coefficients of $t^6\xi^2$. We see that there are four relevant operators, and the number of marginal operators minus the number of conserved currents is equal to eight. As we expect a single $U(1)$ flavor symmetry to be preserved everywhere on the conformal manifold, this gives a conformal manifold of the expected dimension,
\be
\dim_\C\cM_{\rm CFT}= 8+1 = 4g-3~.
\ee
There is an interesting nearly marginal operator which appears here, contributing $\xi^{-1}$ to the index. This is an M2-brane operator that in this special case of genus three is fairly light.

We have checked the other possible genus three constructions that are expected to lead to the same fixed point to verify that they match at leading order. We show later in this section that this can be proven exactly using identities on elliptic gamma functions, while in Section \ref{sec:TQFT} we establish duality for general rank.

The $z=0$ theory is special and has $\ve=0$. This is the partner theory to the corresponding $\cN=2$ theory that enjoys the famous central charge ratio of $27/32$ and was studied in \cite{Benini:2009mz}. While the index of this theory can be computed by a simple specialization of the $\cN=2$ index \cite{Gadde:2010en}, using the construction of \cite{Bah:2012dg} allows us to retain information about the flavor fugacity for $F$. We have computed the resulting index for a number of UV quivers, and the leading terms in the index are given by
\be
\cI[\cC_{3,0};z=0] = 7 + 2\chi_{adj}(\xi)-4\chi_\square(\xi)\chi_{\square}(y)~.
\ee
The dependence on $\xi$ organizes into characters of $SU(2)$ because there are points on the conformal manifold of this theory where $U(1)_F$ is enhanced to $SU(2)_F$ symmetry. For this theory, there is no conserved current at a generic point on the conformal manifold, and so the index counts precisely the number of exactly marginal operators, of which there are $13 = 7 + 2\dim(adj)$. This is one larger than the usual $6g-6$ because the M2 brane operator is exactly marginal at rank one and genus three. Finally, there are no relevant (chiral) operators, which is as expected.

\begin{figure}[t]
\centering
\includegraphics[width=4.25in]{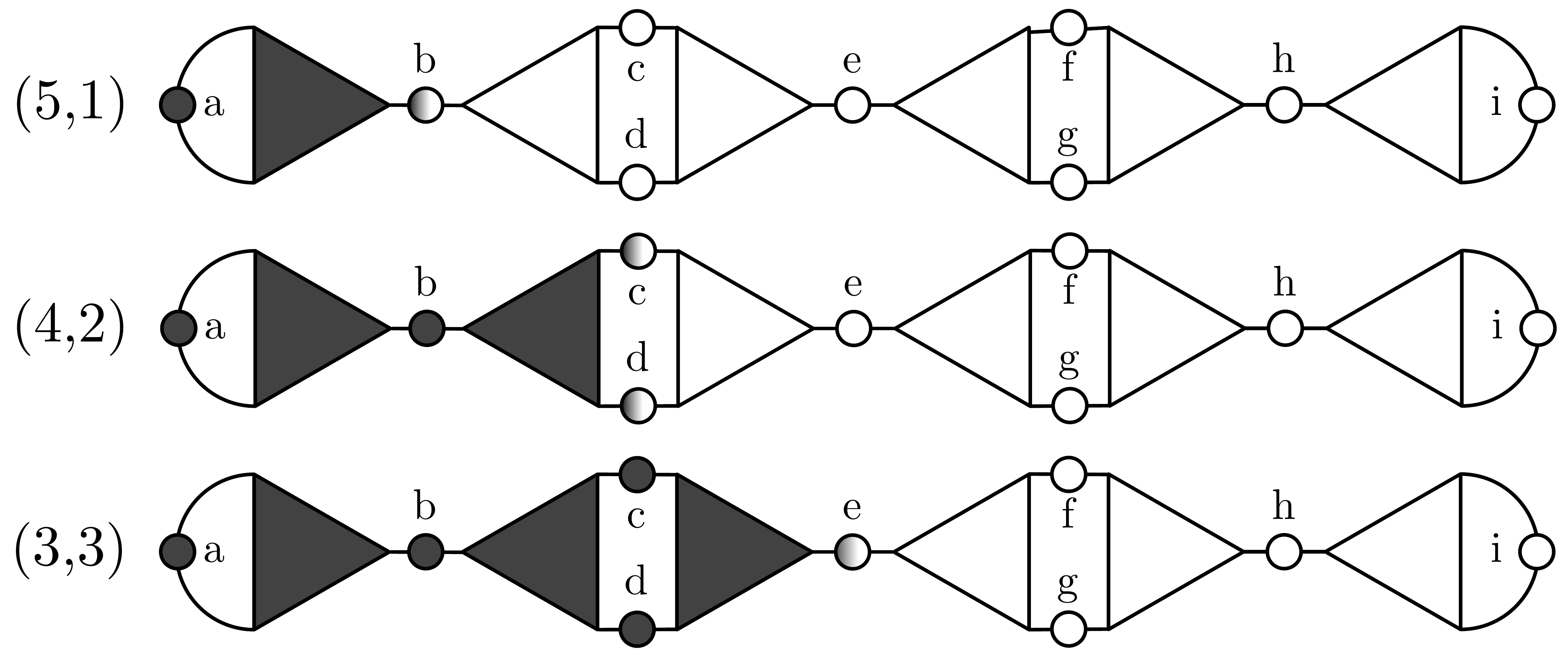}
\caption{Representatives constructions of all genus four generalized quivers with $|z|<1$.}
\label{fig:genus4}
\end{figure}

\subsubsection*{Genus four}
We have also computed the index for genus four theories at all intermediate levels (see Figure \ref{fig:genus4}) to high enough order to count relevant and marginal operators. The terms that can contribute are all at order $t^6$, and are given by
\be\ba
\cI[\cC_{4,0};z=0]&=\(9+3\chi_{adj}(\xi)-6\chi_{\square}(\xi)\chi_{\square}(y)\)t^6+\cO(t^9)~,\\
\cI[\cC_{4,0};z=\tfrac 13]&=(12+5\xi^2+\xi^{-2}-(7\xi+5\xi^{-1})\chi_{\square}(y))t^6+\cO(t^9)~,\\
\cI[\cC_{4,0};z=\tfrac 23]&=(12+7\xi^2-\xi^{-2}-(8\xi+4\xi^{-1})\chi_{\square}(y))t^6+\cO(t^9)~.
\ea\ee
For $z=\tfrac 12,\tfrac 23$ we have $\ve>0$, and so the exactly marginal operators appear only in the constant terms. For $z=0$, however, $\ve=0$ and the number of marginal operators is enhanced and includes operators that are charged under $F$. We can therefore read off the dimension of the conformal manifold of these theories,
\be
\dim_\C\cM_{\rm CFT}=
\begin{cases}
13 = 12+1 & \mbox{for \;\,} z=\mbox{ $\tfrac 13$, $\tfrac 23$}~,\\
18 & \mbox{for \;\,} z=\mbox{ $0$}~.
\end{cases}
\ee
The number of relevant operators in these theories is also unambiguous, and we find
\be
\#\;{\rm relevant}=
\begin{cases}
0 & \mbox{for \;\,} z=\mbox{ $0$}~,\\
5 & \mbox{for \;\,} z=\mbox{ $\tfrac 13$}~,\\
7 & \mbox{for \;\,} z=\mbox{ $\tfrac 23$}~.\\
\end{cases}
\ee

\subsection{Duality at rank one: $\cN=2$ and $\cN=1$ crossing symmetry} \label{identities}
\begin{figure}\label{basicmoves}
\begin{centering}
{\includegraphics[scale=0.24]{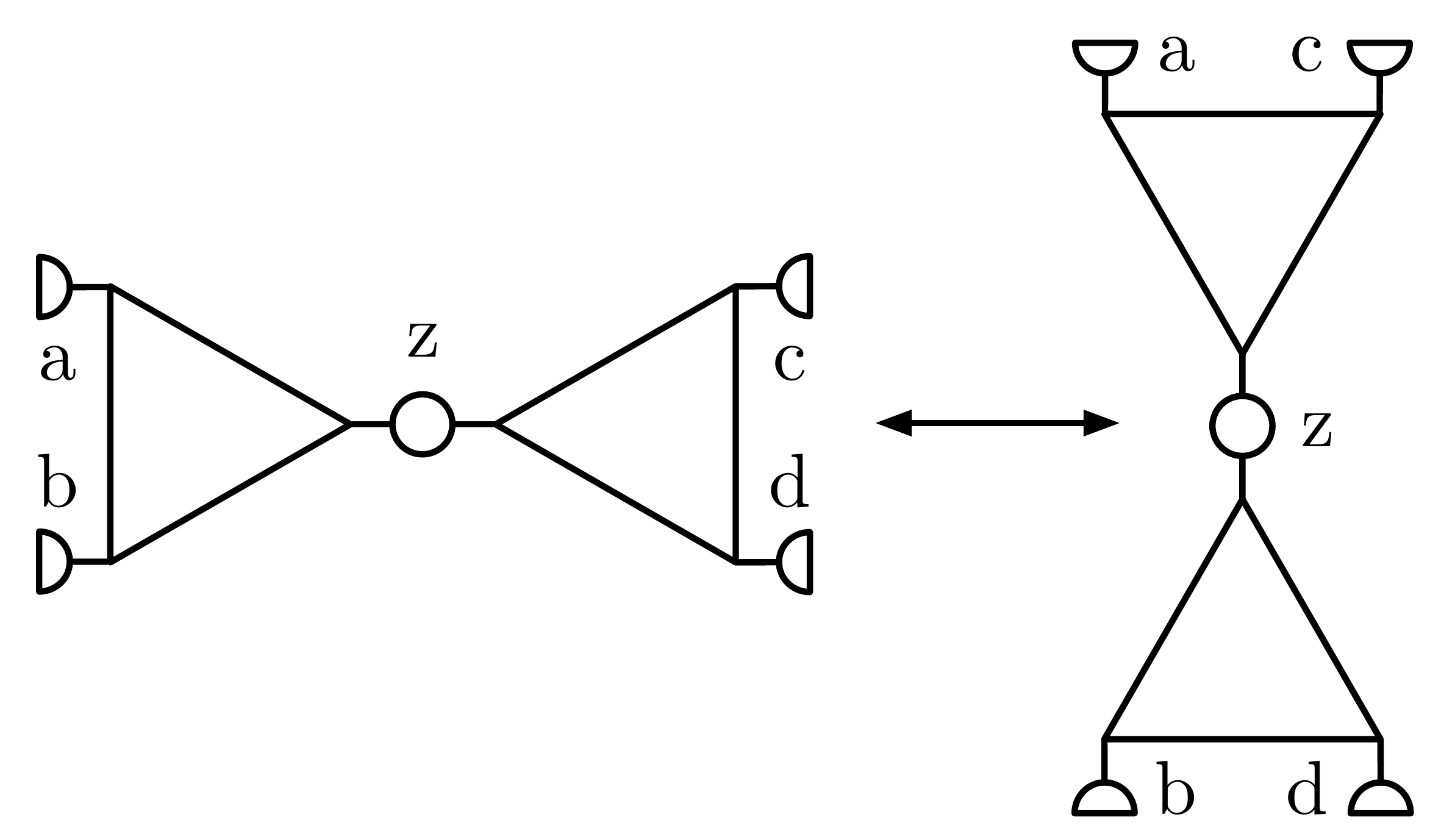}}
$~\quad\quad\quad~$
{\includegraphics[scale=0.24]{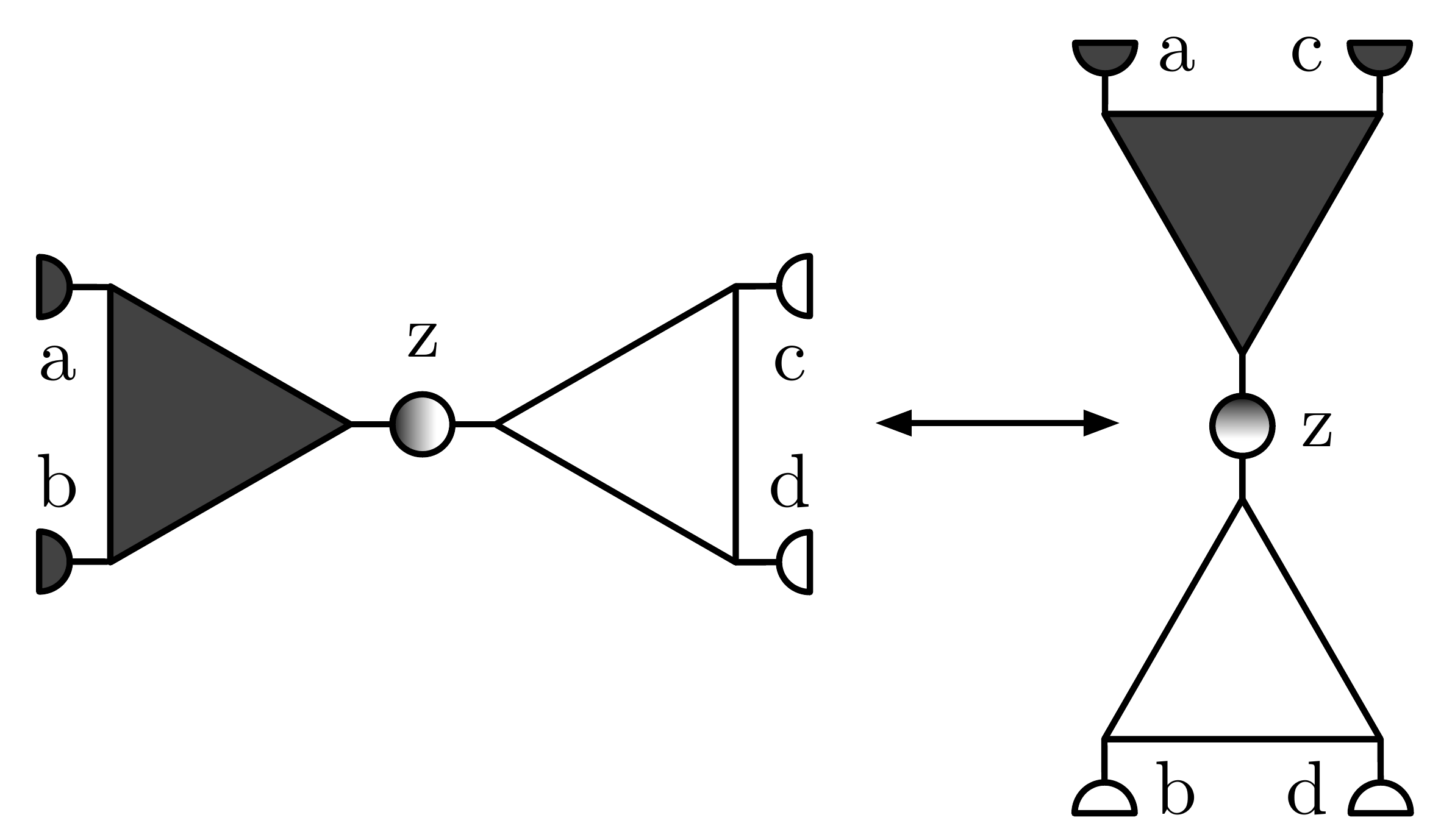}}
\par\end{centering}
\caption{The two local moves that allow arbitrary modifications of a refined pants decomposition within a fixed topological class, demonstrated on ``renormalized'' trinions (indicated by semi-circular external legs). The first figure denotes crossing symmetry at an $\cN=2$ gluing, while the second shows the crossing symmetry at an $\cN=1$ gluing.}
\end{figure}

We can go beyond expansions of the index and understand equivalence of different UV constructions systematically in a way which is analogous to the understanding of $S$-duality for $\cN=2$ theories \cite{Gadde:2009kb}. In doing so, it is helpful to observe that the indices of the vector multiplets listed in \eqref{SU(2)indices} obey a simple relation,
\be
\cI^{\emptyset}_V(a)=\( \cI^+_V(a) \cI^-_V(a)\)^{\frac 12}.
\ee
This enables us to define ``renormalized'' indices of the $T_2$ theories,
\ben
{\tilde \cI}_{T_2}^+(a_1,a_2,a_3)& := &\cI_{T_2}^+(a_1,a_2,a_3)\( \cI_V^+(a_1) \cI_V^+(a_2)\cI_V^+(a_3)\)^{\frac 12}~,\notag\\
{\tilde \cI}_{T_2}^-(a_1,a_2,a_3)& := &\cI_{T_2}^-(a_1,a_2,a_3)\( \cI_V^-(a_1) \cI_V^-(a_2)\cI_V^-(a_3)\)^{\frac 12}~,
\een
in terms of which the index of the accessible theories can be written in a very simple form. For example, the indices in Equation \eqref{genus3equality} become
\be\ba
\cI_A(\xi;t,y)&=\oint[da\ldots dg]{\tilde \cI}_{T_2}^{-}(a,a,b){\tilde \cI}_{T_2}^{+}(b,c,e){\tilde\cI}_{T_2}^{+}(c,e,f){\tilde\cI}_{T_2}^{+}(f,g,g)~,\\
\cI_B(\xi;t,y)&=\oint[da\ldots dg]{\tilde \cI}_{T_2}^{+}(a,a,b){\tilde \cI}_{T_2}^{-}(b,c,e){\tilde\cI}_{T_2}^{+}(c,e,f){\tilde\cI}_{T_2}^{+}(f,g,g)~.
\ea\ee
For any Riemann surface with a refined pants decomposition, one can write the integrand by multiplying the contribution ${\tilde \cI}^\pm_{T_2}$ from $\pm$ type $T_N$ theories. It is clear that the equality of the index for \emph{all} UV constructions of the same topological type will follow from invariance under two elementary local moves illustrated in Figure \ref{basicmoves}.
They amount to crossing symmetry at an $\cN=2$ node (which is equivalent to $\cN=2$ $S$-duality) and crossing symmetry at an $\cN=1$ node. This invariance amounts to the following identities on integrals of the renormalized indices,
\ben
\oint [dz] {\tilde \cI}_{T_2}^+(a,b,z) {\tilde \cI}_{T_2}^+(z,c,d)&=&\oint [dz]{\tilde \cI}_{T_2}^+(a,c,z){\tilde \cI}_{T_2}^+(z,b,d)~,\\
\oint [dz] {\tilde \cI}_{T_2}^+(a,b,z) {\tilde \cI}_{T_2}^-(z,c,d)&=&\oint [dz]{\tilde \cI}_{T_2}^+(a,c,z){\tilde \cI}_{T_2}^-(z,b,d)~.
\een

Formulated in terms of elliptic gamma functions, these identities can be expressed quite explicitly. First, let us define
\be\begin{split}
G^{\cN=2}(a,b;c,d)&:=\oint\frac{dz}{2\pi iz}\frac{\Gamma(\sqrt{pq} \xi z^{\pm2})}{\Gamma(z^{\pm2})}\Gamma((pq)^{\frac 14}\xi^{-\frac 12} a^\pm b^\pm z^\pm)\Gamma((pq)^{\frac 14}\xi^{-\frac 12} z^\pm c^\pm d^\pm)~,\\
G^{\cN=1}(a,b;c,d)&:=\oint\frac{dz}{2\pi iz}\frac{1}{\Gamma(z^{\pm2})}\Gamma((pq)^{\frac 14}\xi^{\frac 12} a^\pm b^\pm z^\pm)\Gamma((pq)^{\frac 14}\xi^{-\frac 12} z^\pm c^\pm d^\pm)~.
\end{split}\ee
Invariance under the three basic moves then requires that the following identities hold,
\ben
\mbox{$\cN=2$ crossing:\qquad}G^{\cN=2}(a,b;c,d)&=&G^{\cN=2}(a,c;b,d)\label{N=2crossing}~,\\
\mbox{$\cN=1$ crossing:\qquad}G^{\cN=1}(a,b;c,d)&=&\frac{\Gamma((pq)^{\frac 12} \xi b^{\pm2})}{\Gamma((pq)^{\frac 12} \xi c^{\pm2})}G^{\cN=1}(a,c;b,d)\label{N=1crossing}~.
\een
The $\cN=2$ identity \eqref{N=2crossing} has been proved in \cite{Bult:2009aa}, and amounts to $S$-duality for rank one $\cN=2$ theories of class $\cS$ {\cite{Gadde:2009kb}. Remarkably, the identity \eqref{N=1crossing} is also known to be true --- it follows from $E_7$ Weyl symmetry of the elliptic beta integral considered in \cite{Spiridonov:2003fk,Spiridonov:2008zr}.\footnote{We thank Wenbin Yan for pointing out this consequence.} The existence of these identities amounts to a proof of equality for the index of all the rank one accessible theories in a given topological class.

Some of the formal aspects of the analysis above apply equally well to the higher rank theories. However, we will bypass this by introducing the TQFT structure of the index in order to determine the higher rank index. This is the subject to which we now turn.

\section{Topological quantum field theory and the class $\cS$ index}\label{sec:TQFT}

The superconformal index for $\cN=2$ fixed points admits an efficient formulation in the language of two-dimensional topological quantum field theory \cite{Gadde:2009kb}. This interpretation follows from the identification of the conformal manifold of these theories with the moduli space of the UV curve, and the approach plays an important role in the solution for the index of higher rank, non-Lagrangian theories \cite{Gaiotto:2012xa}. This section begins with a review of the TQFT that computes this $\cN=2$ index. After reminding the reader of the ingredients that enter into that construction, we extend the construction in order to compute the index of accessible $\cN=1$ fixed points. This leads to a formal description of the TQFT for $\cN=1$ fixed points, which suggests a natural conjecture for the inaccessible fixed points as well. 

The fugacity conventions used in this section match those of \cite{Gadde:2011uv}. This means that the $\cN=2$ indices described here are defined by
\be\label{N=2fugs-tqft}
\cI_{\cN=2}(a;p,q,\t)=\Tr(-1)^Fp^{\cJ_{12}-r}q^{\cJ_{34}-r}\t^{R+r}\prod_{i}a_i^{f_i}~,
\ee
with $R$ the cartan of $SU(2)_R$ and $r$ the generator of $U(1)_r$.\footnote{The index here is defined with respect to the supercharge $\wt\cQ_{1\dot{-}}$, which has charges $\(\tfrac 12,-\tfrac 12\)$ under $(R,r)$.} The $\cN=1$ indices are as defined in Equation \eqref{simpleindex},
\be
\cI_{\cN=1}(a,\xi;p,q)=\Tr(-1)^Fp^{\cJ_{12}-R_0}q^{\cJ_{34}-R_0}\xi^F\prod_{i}a_i^{f_i}~.
\ee
When the $\cN=1$ index is evaluated on $\cN=2$ theories of ``$\pm$'' type (or constituents of larger theories), the $R$-symmetries are related by $R_0 =  R + r$ and $F=\pm(R-r)$. The $\cN=2$ and $\cN=1$ indices are then related by
\be
\cI^{\pm}(a,\xi;p,q)=\cI_{\cN=2}(a;p,q,\xi^{\pm1}(pq)^{\frac 12})~.
\ee
Note that $R_0$ is the $R$-symmetry that is preserved when $\cN=2$ supersymmetry is broken to $\cN=1$ by introducing a superpotential mass term for the adjoint-valued chiral multiplet in the vector multiplets.

\subsection{The $\cN=2$ index}

The index of an $\cN=2$ fixed point is equal to a correlation function for a two-dimensional topological quantum field theory on the curve $\cC_{g,h}$ with appropriate operators inserted at the punctures. The three-point functions of this TQFT for a canonical set of operators are equal to the index of the trinion theory $T_N$. When $N=2$, this is simply the free trifundamental chiral multiplet index in \eqref{SU(2)indices}, while for $N=3$ this is the index of the $E_6$ SCFT, which has been computed in \cite{Gadde:2010te}. For $N>3$, a procedure for generating all of such indices in principle has been given in \cite{Gaiotto:2012xa}, while certain limits of the index are known in closed form, \cf\ \cite{Gadde:2011uv}.

The index of a trinion theory $T_N$ can be written in the form
\be\label{trinindex}
\cI_N[\cC_{0,3};z=0](a,b,c;p,q,\t)=\sum_{\alpha}C_\alpha(p,q,\t)\,\psi^\alpha(a;p,q,\t)\,\psi^\alpha(b;p,q,\t)\,\psi^\alpha(c;p,q,\t)~,
\ee
where $\alpha$ runs over all irreducible representations of $SU(N)$. This representation of the index makes evident many of the salient features of the index TQFT. Firstly, the expression in \eqref{trinindex} is a combination of two elementary parts: the \emph{structure constants} $C_\alpha(p,q,\t)$ and the \emph{wave-functions} $\psi^\alpha(a;p,q,\t)$. The structure constants are interpreted as the partition function of the TQFT on a sphere with three $S^1$ boundary components, where a state $\alpha$ has been imposed on each component. The Hilbert space of the TQFT is spanned by the irreps of $SU(N)$, and in this basis the structure constants are diagonal. The structure constants for a fixed choice of external state are functions of only the superconformal fugacities and may in principle be determined by a careful application of $S$-duality (\cf\ \cite{Gaiotto:2012xa}). 

The wave-functions are functions of the flavor fugacities that enter into the definition of the index, and they describe an operator insertion at a puncture --- or alternatively, an insertion of a non-normalizable state at each boundary component which is a delta function in the fugacity basis. They are orthonormal with respect to the measure defined by the index of an $\cN=2$ vector multiplet,
\be\label{orthonormal}
\oint[da]\,\cI_V^{\cN=2}(a)\psi^\alpha(a)\psi^\beta(a)=\delta^{\alpha\beta}~,
\ee
where again the measure $\oint[da]$ denotes an integral over the maximal torus of $SU(N)$ with the Haar measure. From the point of view of the four-dimensional gauge theory, the fugacity basis is the natural one, and gluing is accomplished by gauging a diagonal subgroup of two $SU(N)$ global symmetries associated to the glued legs. The effect on the index is to integrate over fugacities of the gauged symmetry with an insertion of the $\cN=2$ vector multiplet index, and the relation \eqref{orthonormal} is responsible for the index being related to a TQFT which is simple in the representation basis. The wave functions can be approximated to arbitrary order using the approach of \cite{Gaiotto:2012xa}. 

In terms of these constituent functions (which may themselves be complicated and difficult to determine), it is a simple matter to write down a formal expression for the index of the theory at genus $g$ with $h$ punctures,
\be\label{N=2indextop}
\cI[\cC_{g,h}]=\sum_\alpha \left((C_\alpha)^{2g+h-2}\prod_{i=1}^h\psi^\alpha(a_i)\right)~.
\ee
When written in this way as a TQFT correlator, $S$-duality is manifest. Indeed, this was the motivation behind the original search for a TQFT definition of the index. To go beyond this formal expression for $N>3$ requires implementing of the methods outlined in \cite{Gaiotto:2012xa} for determining the higher rank $T_N$ structure constants and wavefunctions.

\subsection{The $\cN=1$ index of all accessible fixed points}\label{subsec:TQFT}

In the language of the $\cN=2$ TQFT, it is straightforward to describe the index of the accessible $\cN=1$ fixed points. This generalizes the procedure outlined in Section \ref{sec:SU(2)index} for rank one theories. As in that case, it is not \emph{prima facie} obvious is that the resulting partition functions will be a correlators in a modified TQFT. In particular, we shall see that this property of the index requires that certain identities be satisfied by the wavefunctions.

In the $\cN=1$ setup, there are two distinct three-point functions, corresponding to the choice of a $(1,0)$ or $(0,1)$ type of trinion. As in the rank one case, these differ only in the identification of the fugacities of the $T_N$ index with the globally defined fugacities of the theory. We independently decompose the two three-point functions into structure constants and wavefunctions, leading to a doubling of both types of objects,
\be\label{s3pt}
C_\alpha^{\pm}\(\xi;p,q\) = C_\alpha\(p,q,\xi^{\pm1}(pq)^{\frac 12}\)~,
\ee
\be\label{swave}
\psi^{\alpha}_{\pm}\(a,\xi;p,q\)=\; \psi^\alpha\(a;p,q,\xi^{\pm1}(pq)^{\frac 12}\)~.
\ee
In stitching together geometries as described in Section \ref{sec:b3wrev}, a ``$+$'' trinion appears in the index with ``$+$'' three-point functions, while a ``$-$'' trinion will appear with ``$-$'' three point functions. 

In addition, there are now three different types of gluing in the fugacity basis, corresponding to $\cN=2$ gauging of type $\pm$ or $\cN=1$ gauging. We define the appropriate measures for the different types of gauging as in Section \ref{sec:SU(2)index},
\be\label{sprop}
\cI_V^{\emptyset,\pm}\(a,\xi;p,q\)=\cI_V\(a;p,q,\xi^{s}(pq)^{\frac 12}\)~,\qquad s=\emptyset,\pm1~.
\ee
We have leveraged the fact that specialization the fugacities of an $\cN=2$ vector multiplet as is done for $s=0$ automatically leads to the index of an $\cN=1$ vector multiplet, which is the appropriate measure for gluing together opposite types of trinions.

The simplicity of the final expression for the $\cN=2$ index in \eqref{N=2indextop} is a consequence of the orthonormality of the wave functions under the measure defined by the $\cN=2$ vector multiplet. It is trivial to observe that for both types of $\cN=2$ gaugings here, the orthonormality property carries over since the only modification is a uniform redefinition of fugacities for all three functions involved. It is less obvious what will happen at $\cN=1$ gaugings. Let us consider this case. The integral for the $\cN=1$ gauge group is given by
\be\label{N=1glue}
\oint [da]\,\cI_V^{\emptyset}(a)\,\psi^\alpha_{+}(a)\,\psi^\beta_{-}(a)~,
\ee
Without additional knowledge of the wavefunctions we would be at an impasse. However, it was found in \cite{Gaiotto:2012xa} that the wavefunctions obey an important identity,
\be
\frac{1}{\prod_{i\neq j}\Gamma(\t \frac{a_i}{a_j};p,q)}\psi^\alpha(a;p,q,\t)=\psi^\alpha(a;p,q,\tfrac{pq}{\t})~.
\ee
In the present context, this relation amounts to a map between the two types of wavefunctions,
\be\label{wavemap}
\ba
\psi_+^\alpha(a)&=\psi_-^\alpha(a)\prod_{i\neq j}\Gamma\((pq)^{\frac12}\xi \frac{a_i}{a_j};p,q\)~,\\
\psi_-^\alpha(a)&=\psi_+^\alpha(a)\prod_{i\neq j}\Gamma\((pq)^{\frac12}\xi^{-1} \frac{a_i}{a_j};p,q\)~.\\
\ea
\ee
These two transformations are self-consistent because of the identity for elliptic gamma functions,
$\Gamma(z;p,q)\Gamma(\tfrac{pq}{z};p,q)=1$.
Remarkably, this identity precisely guarantees that the opposite types of wavefunctions are mutually orthonormal with respect to the $\cN=1$ measure,
\be
\psi^\alpha_{+}(a)\psi^\alpha_{-}(a)\cI_V^{\emptyset}(a)=\psi^\alpha_{+}(a)\psi^\alpha_{+}(a)\cI_V^{+}(a)=\psi^\alpha_{-}(a)\psi^\alpha_{-}(a)\cI_V^{-}(a)~.
\ee 
This result immediately yields the index for a constructible theory with $h_1$ external legs of ``$+$'' type, $h_2$ external legs of ``$-$'' type, $\ell_1$ copies of $T_N^+$, and $\ell_2$ copies of $T_N^-$:
\be\label{constructibleindex}
\cI\[\cC_{g,h_1,h_2}^{\ell_1,\ell_2}\]=\sum_\alpha\(C_\alpha^+(\xi;p,q)^{\ell_1}C_\alpha^-(\xi;p,q)^{\ell_2}\prod_{i=1}^{h_1}\psi^\alpha_+(a_i,\xi;p,q)\prod_{j=1}^{h_2}\psi^\alpha_-(b_j,\xi;p,q)\)~.
\ee
Though this expression is abstract, the result establishes that any two constructions of an accessible $\cN=1$ fixed point of the same topological type will yield a theory with the same index, thus generalizing the proof of universality for the index from Section \ref{sec:SU(2)index}. This is a strong check that all such UV constructions do indeed belong to the same universality class.

\subsection{TQFT and inaccessible fixed points}
%
%
%
%
\newcommand{\prop}{     \includegraphics[height=3.5ex]{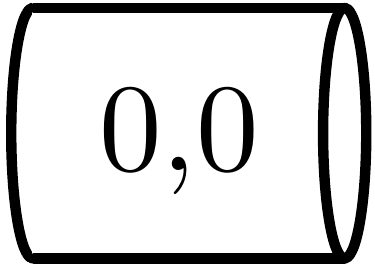}}
\newcommand{\capleft}{  \includegraphics[height=3.5ex]{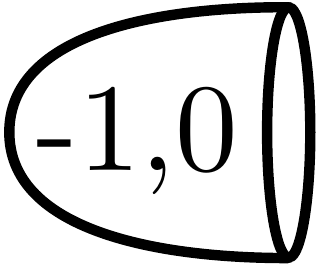}}
\newcommand{\capright}{ \includegraphics[height=3.5ex]{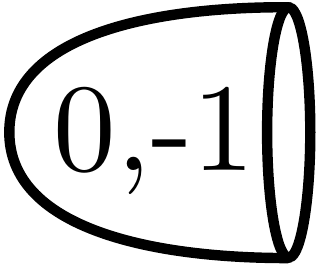}}
\newcommand{\proppm}{   \includegraphics[height=3.5ex]{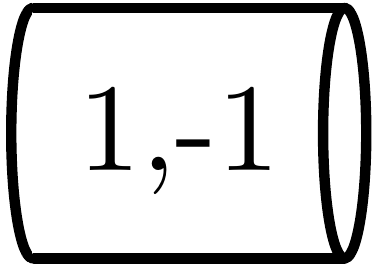}}
\newcommand{\propmp}{   \includegraphics[height=3.5ex]{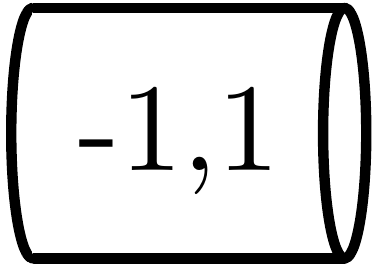}}
\newcommand{\trinleft}{ \includegraphics[height=6ex]{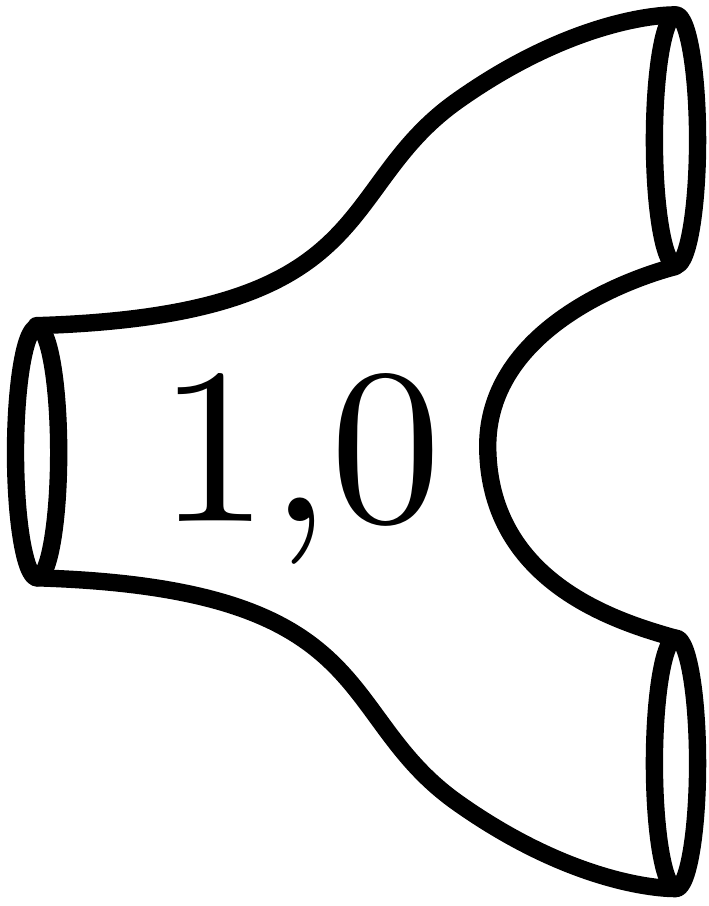}}
\newcommand{\trinright}{\includegraphics[height=6ex]{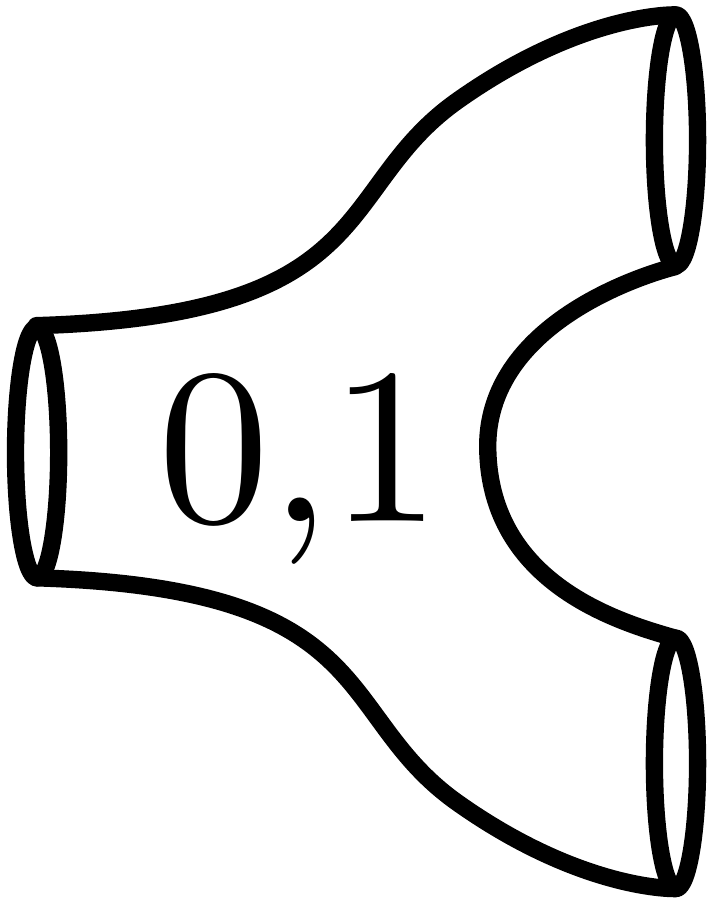}}
%
%
%
%
The construction described above does not quite fit into the standard framework of a two-dimensional topological quantum field theory -- there is extra discrete data associated to the line bundles over every pair of pants. Such a construction has appeared previously in the description of the local Gromov-Witten theory of curves \cite{Bryan:2004iq}. Technically speaking, the effect of keeping track of the line bundles is to replace the two-dimensional cobordism category that is relevant to the Atiyah-Segal formulation of two-dimensional TQFT with the category $2{\bf Cob}^{L,K\otimes L^*}$ of 2-cobordisms endowed with line bundles $L$ and $K\otimes L^*$ (trivialized over the boundaries), with $K$ the canonical bundle of the cobordism.\footnote{This is a specialization of the category used in \cite{Bryan:2004iq} to the case where only ``Calabi-Yau cobordisms'' appear, \ie, those cobordisms over which the determinant of the line bundles is the canonical bundle. While local Gromov-Witten invariants can be formulated without this constraint, it doesn't appear to be possible with the index.}
For the accessible fixed points, this is an overly technical description of what was just discussed in the previous section -- namely that there are two kinds of three-point functions associated with different line bundles over the trinion. However, if the full TQFT defined as such is the correct description for the indices of these $\cN=1$ fixed points, then this implies an expression for the inaccessible fixed points as well.

So far, the elements of the TQFT which are defined are the propagator and three-point functions for trinions of type $(1,0)$ and $(0,1)$. In the TQFT, these evaluate to
\be\ba\label{generators}
Z_{\alpha\beta\gamma}\bigg(\!\!\raisebox{-11pt}{\trinleft}\;\bigg)&=C^+_\alpha(\xi;p,q)\delta_{\alpha\beta}\delta_{\beta\gamma}~,\\
Z_{\alpha\beta\gamma}\bigg(\raisebox{-11pt}{\trinright}\;\bigg)&=C^-_\alpha(\xi;p,q)\delta_{\alpha\beta}\delta_{\beta\gamma}~,\\
Z_{\alpha\beta}\bigg(\!\!\raisebox{-6pt}{\prop}\;\bigg)&=\delta_{\alpha\beta}~.
\ea\ee
These are actually sufficient to determine all Riemann surface topologies with all (admissible) choices of line bundles. In particular the ``cap'' amplitudes at level $(-1,0)$ and $(0,-1)$ are completely fixed by the requirement that they trivialize their respective three-point functions to produce the propagator as shown on the left side of Figure \ref{fig:gluedprop}. The result is simple,
\begin{figure}
\begin{center}$
\begin{array}{cc}
\includegraphics[width=2in]{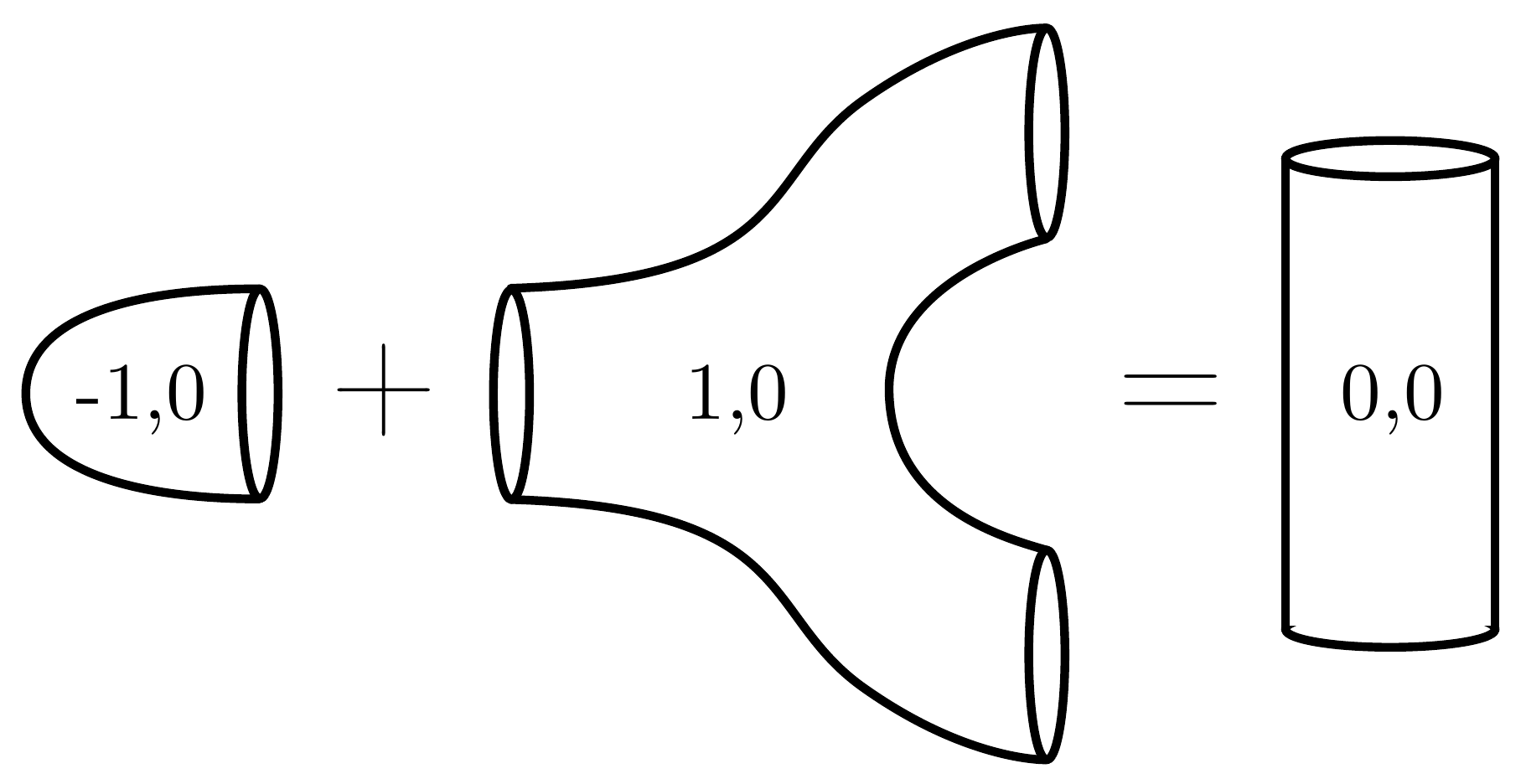}\qquad & \qquad
\includegraphics[width=2in]{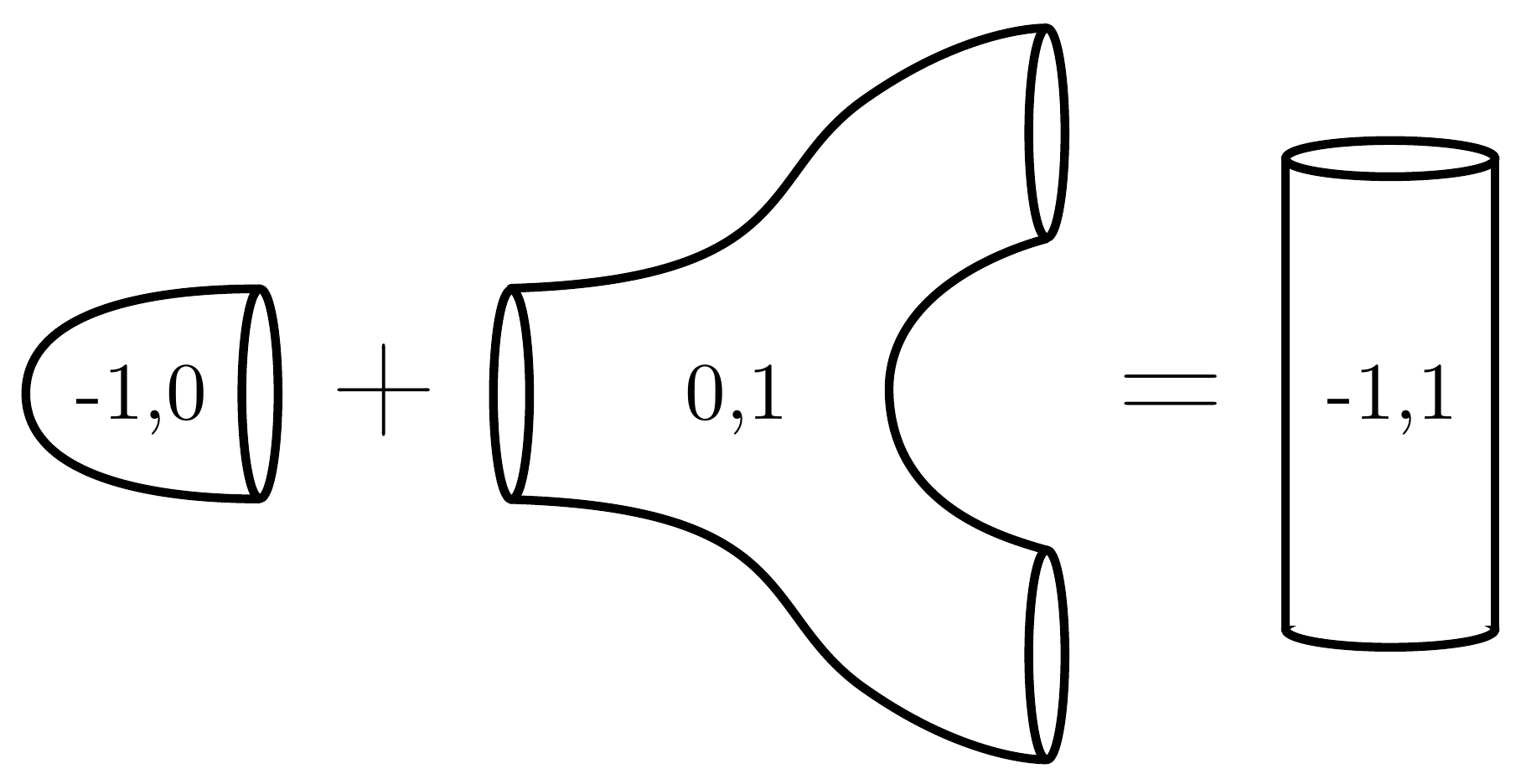}
\label{fig:gluedprop}
\end{array}$
\end{center}
\caption{[left] A propagator is obtained from the trinion and cap amplitudes. This determines the cap amplitude to be a simple inverse of the trinion amplitude in the diagonal basis. [right] Gluing a cap and trinion with a different choice of levels leads to a propagator that shifts the overall level of the cobordism.}
\end{figure}

\be\label{cap-amplitude}
Z_\alpha\Big(\!\!\raisebox{-5.5pt}{\capleft}\,\Big)=\frac{1}{C_{\alpha}^+(\xi;p,q)}~,\qquad\qquad\;\;
Z_\alpha\Big(\!\!\raisebox{-5.5pt}{\capright}\,\Big)=\frac{1}{C_{\alpha}^-(\xi;p,q)}~.\qquad
\ee
By applying the same cap amplitude to the opposite type of three-point function, we obtain a new propagator that shifts the levels of the two line bundles by $\pm1$,
\be\label{shift-prop}
Z_{\alpha\beta}\Big(\!\raisebox{-5.5pt}{\propmp}\,\Big)=\frac{C_{\alpha}^-(\xi;p,q)}{C_{\alpha}^+(\xi;p,q)}\delta_{\alpha\beta}~,\qquad
Z_{\alpha\beta}\Big(\!\raisebox{-5.5pt}{\proppm}\,\Big)=\frac{C_{\alpha}^+(\xi;p,q)}{C_{\alpha}^-(\xi;p,q)}\delta_{\alpha\beta}~.
\ee

These amplitudes are all extremely simple in the representation basis, leading to a simple structure for the higher-level indices. In particular, it is easy to see that by inserting factors of the level-shifting propagator, one can construct a geometry with arbitrary (allowed) degrees $\ell_1$ and $\ell_2$. By additionally allowing the insertion of both ``$+$'' and ``$-$'' type operators on the curve, we find that the general form for the $\cN=1$ index is still given by \eqref{constructibleindex}, but where the levels $\ell_1,\ell_2$ are no longer constrained to be non-negative,
\be\label{inaccessibleindex}
\cI\[\cC_{g,h_1,h_2}^{\ell_1,\ell_2}\]=\sum_\alpha \( C_\alpha^+(\xi;p,q)^{\ell_1}C_\alpha^-(\xi;p,q)^{\ell_2}\prod_{i=1}^{h_1}\psi^\alpha(a_i,\xi;p,q)\prod_{j=1}^{h_2}\psi^\alpha(b_j,\xi;p,q)\)~.
\ee
Thus we are led to indices for all $\cN=1$ fixed points of class $\cS$, including exotic constructions like the theories with a closed sphere as the UV curve. These results are fairly formal, however, as the elementary structure constants and wave functions are not known in general. We now turn to consider the behavior of the $\cN=1$ index when fugacities are specialized in a way that simplifies these functions.

\section{Simplifying limits}\label{sec:simplelimits}

There exist a variety limits of the $\cN=2$ index that preserve additional supersymmetry, and so count only a subset of the usual protected states. At the level of a trace formula, these limits come about in two different ways: either by specializing the fugacities of the index in such a way that the resulting trace can be interpreted as an index for two different supercharges, or by taking a limit where the fugacity of a strictly non-negative operator becomes zero, effectively restrict the range of the trace. All such limits which exist for general $\cN=2$ SCFTs were described in \cite{Gadde:2011uv}.\footnote{It should be noted that for the $\cN=1$ index, there is a single limit that superficially leads to an enhancement of supersymmetry: by setting $q=1$ in \eqref{indexdef}, the same trace can be interpreted as both a left- and right-handed index. This is a singular limit, with any non-zero contribution to the index diverging. While it is imaginable that some useful information could be extracted from residues of this singular point, we will not pursue the subject further here.}
In describing these limits, it is useful to introduce a reparameterization of the fugacities 
\be\label{positivefugacities}
p=\rho\tau~,\quad q=\sigma\tau~,\quad\t=\tau^2~,
\ee
which appear in the $\cN=2$ index with only positive powers, and so one can set them to zero without any second thoughts. In terms of these fugacities, the relation between ``$+$'' type and ``$-$'' constituents is
\be
\tilde\rho=\tau\sqrt{\frac{\rho}{\sigma}}~,\quad\tilde\sigma=\tau\sqrt{\frac{\sigma}{\rho}}~,\quad\tilde\tau=\sqrt{\rho\sigma}~.
\ee%

By taking certain limits of the $\cN=1$ index for the theories studied in the present work, we can realize some of the enhanced limits for the $\cN=2$ building blocks, leading to known simplifications for the constituent indices. From the point of view of superconformal representation theory, these limits are fairly unnatural as they mix the flavor fugacities $\xi$ with superconformal fugacities in a way which is only meaningful due to the $\cN=2$ supersymmetry present in the UV construction. Indeed, the approach of this section can be applied to any $\cN=1$ theory constructed out of $\cN=2$ supersymmetric constituent theories in such a way as to preserve a canonical flavor symmetry such as $F$ in our theories. We find that the special limits of \cite{Gadde:2011uv} are effectively paired in the $\cN=1$ constructions. Because these simplifying limits lead to tractable expressions for the TQFT structure constants, we will be able to perform more explicit calculations for the higher rank theories and extract useful information about light operators.

\subsection*{Macdonald limit}

The Macdonald index of an $\cN=2$ theory is defined by the specialization of fugacities $\rho \to0$, $\sigma,\tau$ fixed. It is clear that for a limit of the $\cN=1$ index that reproduces the Macdonald index for a set of $\cN=2$ constituents, the index of the opposite type of constituents will be singular ($\tilde\sigma\to\infty$). There is thus no good relative of the Macdonald index for the $\cN=1$ fixed points. We now observe that the situation is markedly improved for the other limits.

\subsection*{Mixed Schur index}

The \emph{Schur index} is defined by the specialization $\sigma=\tau$. The fugacity $\rho$ drops out of the Schur index of an $\cN=2$ theory. When this limit is applied to ``$+$'' type constituents, the effect on the ``$-$'' type constituents will be $\tilde\rho=\tilde\tau$. Thus the result is a complementary Schur index for which $\tilde\sigma$ is irrelevant. This specialization is implemented at the level of the $\cN=1$ index by fixing $\xi = \sqrt{q/p}$, which is a perfectly acceptable specialization, although it would be unusual for a generic $\cN=1$ theory. Nonetheless, we see that two different simplified limits of the $\cN=2$ index are paired together by the $\cN=1$ fixed points, making this limit of the index quite computable.

The $\cN=2$ TQFT in the Schur limit is known to reduce to the zero area limit of $q$-deformed two-dimensional Yang-Mills theory. The three point functions and wave-functions are known for the $T_k$ theory,
\be
C^\alpha(p)=\frac{(q;q)}{\prod_{j=1}^{k-1}(1-q^j)^{k-j}}\frac{1}{\dim_q\alpha}~,\qquad \psi^\alpha(a)=\frac{\chi^\alpha(a)}{\[\cI_V^{\cN=2}(a)\]^{\frac 12}}~,
\ee
Thus the specialization of the general index \eqref{inaccessibleindex} is given by
\be
\cI = \frac{
(p;p)^{\ell_1}
(q;q)^{\ell_2}
}
{
\prod_{i=1}^{k-1}(1-p^i)^{\ell_1(k-i)}
\prod_{j=1}^{k-1}(1-q^j)^{\ell_2(k-j)}
}\sum_\alpha\left(\frac{1}{(\dim_p\alpha)^{\ell_1}(\dim_q\alpha)^{\ell_2}}\right)~.
\ee
The physical content of this limit is somewhat obscure. The specialization mixes the fugacities for $SU(2)_2$ with the canonical flavor symmetry $F$. We cannot extract net degeneracies from the index defined as such. For example, simple relations such as
\be
1+\chi_{adj}(y)-(\xi+\xi^{-1})\chi_\square(y) \xrightarrow[\xi=y]{} 0~,
\ee
allow for new cancellations in the specialized case, destroying much of the information content in the index. 

\subsection*{Hall-Littlewood/Coulomb index}

The final set of special limits are those dubbed the \emph{Hall-Littlewood} (HL) \emph{index} and \emph{Coulomb index} in \cite{Gadde:2011uv}. The HL index appears in the limit $\rho\to0$, $\sigma\to0$, $\tau$ fixed. The complementary limit is given by $\tilde\tau\to0$, $\tilde\rho$ and $\tilde\sigma$ fixed, which is the Coulomb index. This combined limit is obtained in the $\cN=1$ fugacities by taking
\be
\xi := \frac{w}{t^3}~,\quad t\to0~.
\ee
Notice that because the limits act differently on the two types of trinions, the resulting index will not be invariant under the exchange of the fibers in the Calabi-Yau geometry, $\ell_1\leftrightarrow\ell_2$. Indeed, the index which will produce the same result for the flipped geometry will arise from setting $\xi = \frac{t^3}{w}$.

Unlike the mixed Schur index, this limit can understood to be physically meaningful at the level of superconformal representation theory. The contribution of a short multiplet to the $\cN=1$ index in this limit is given by
\be
\cI_{[\tilde{r},j_{2}]_{+}}^{{\tt L}}=-\cI_{[\tilde{r},j_{2}]_{-}}^{{\tt L}}=\lim_{t\to0}(-1)^{2j_{2}+1}\frac{t^{3(\tilde{r}_0+2-F)}\chi_{j_{2}}(y)w^F}{(1-t^{3}y)(1-t^{3}y^{-1})}~,
\ee
where $\tilde r_0 = 2j_1 - R_0$. In order for this limit to be well-defined, it is necessary that $F < \tilde r_0 + 2$ for all short multiplets that contribute to the index. This is not a necessary condition for a general  $\cN=1$ SCFT with a flavor symmetry, but it is guaranteed for us by the $\cN=2$ structure in the UV and the canonical choice of $F$.

Additionally, it is an empirical fact that the Coulomb index for class $\cS$ theories depends only on $\sigma\rho$, which in this case means that there are no contributions from multiplets with nonzero $j_2$. Thus, the only multiplets counted in this limit are then those which obey the conditions
\be
\tilde{r}_0=F-2~,\qquad j_2=0~,
\ee
and the net degeneracy of these states appears as the coefficient of $w^F$ in the index. In particular, this means that at quadratic order, the index counts operators that were ``marginal'' with respect to the naive R-symmetry $R_0$, but become relevant after using $a$-maximization because of having charge two with respect to $F$. More precisely, this is the case for $\ve>0$, while for $\ve<0$ we must use the opposite index.

While this may sound like a rather special subset of operators to count, a quick examination of the examples in Section \ref{subsec:examples} shows that \emph{all} of the relevant operators in those cases were of precisely this type. Indeed, by examining the structure of the known chiral operators for $T_N$, it appears that in the general case (in particular, when M2 brane operators are not light enough to be relevant), all relevant operators at an $\cN=1$ fixed point must be of this type. Because of the simplicity of the HL and Coulomb indices of $T_N$, we will in fact be able to carry out the computation of the HL/Coulomb limit for a general $\cN=1$ fixed point and find the number of relevant operators. We now describe the calculation and result.

\subsection{Relevant operator counting from the HLC index}
The relevant functions to construct this index are the three-point functions and wave functions in the HL and Coulomb limits. We summarize these functions below:
\be\begin{split}\label{CHLfuncs}
\psi_{+}^{\alpha}(a_{i};p,q,\xi) & \to  \psi_{\rm HL}^{\alpha}(a_{i};\w):=\frac{(1-\w)^{\frac{k-1}{2}}}{\prod_{i\neq j}(1-\w\frac{a_{i}}{a_{j}})}P_{\rm HL}^{\alpha}(a_{i};\w)~,\\
C_{\alpha}^{+}(p,q,\xi) & \to  C_{\alpha}^{\rm HL}(\w):=\frac{\prod_{j=2}^{k}(1-\w^{j})}{(1-\w)^{\frac{k-1}{2}}P_{\rm HL}^{\alpha}(\w^{\frac{k-1}{2}},\w^{\frac{k-3}{2}},\ldots,\w^{-\frac{k-1}{2}};\w)}~,\\
\psi_{-}^{\alpha}(a_{i};p,q,\xi) & \to  \psi_{\rm C}^{\alpha}(a_{i};\w):=\mbox{PE}(-\frac{1}{2}\sum_{j=2}^{k}\w^{j})~,\\
C_{\alpha}^{-}(p,q,\xi) & \to  C_{\alpha}^{\rm C}(\w):=\delta_{\alpha,0}\mbox{PE}(\sum_{j=2}^{k}j\w^{j}-\frac{1}{2}\sum_{j=2}^{k}\w^{j})~.
\end{split}\ee
Here $P_{\rm HL}^{\alpha}$ are the Hall-Littlewood polynomials
\be
P_{\rm HL}^{\alpha}(a_{i};\w):=\left(\prod_{i=0}^{\infty}\prod_{j=1}^{m(i)}\frac{1-\w^{j}}{1-\w}\right)^{-\frac{1}{2}}\sum_{\sigma\in S_{k}}a_{\sigma(1)}^{\alpha_{1}}\ldots a_{\sigma(k)}^{\alpha_{k}}\prod_{i<j}\frac{a_{\sigma(i)}-\w a_{\sigma(j)}}{a_{\sigma(i)}-a_{\sigma(j)}}~,
\ee
where $m(i)$ is the number of rows in the Young diagram $(\alpha_{1},\ldots,\alpha_{k})$ of length $i$. 

The structure constants $C_\alpha^-(p,q,\xi)$ vanish except for the case of the trivial representation $\alpha=0$. The expression \eqref{inaccessibleindex} for the index therefore simplies dramatically,
\be
\cI\left[\cC_{g,h_1,h_2}^{\ell_{1},\ell_{2}}\right]=C_{0}^{HL}(\w)^{\ell_{1}}C_{0}^{C}(\w)^{\ell_{2}}\psi_{\rm C}(w)^{h_2}\prod_{i=1}^{h_1}\psi_{\rm HL}^0(a_i;w)~,\label{HLCindex}
\ee
where the specializations of the functions in \eqref{CHLfuncs} to the trivial representation are given by
\be
\psi_{\rm HL}^{0}(a;w)=\mbox{PE}(-\frac{1}{2}\sum_{j=2}^{k}\w^{j})\mbox{PE}(\w\sum_{i\neq j}\frac{a_{i}}{a_{j}})~,\qquad C_{0}^{\rm HL}(w)=\mbox{PE}(-\frac{1}{2}\sum_{j=2}^{k}w^{j})~.
\ee
We can check the consistency of our specializations by observing that the wavefunctions satisfy the Equation \eqref{wavemap} in this limit,
\be
\psi_{HL}^{0}(a_{i})=\frac{1}{\prod_{i\neq j}(1-\w\frac{a_{i}}{a_{j}})}\psi_{C}^{0}(a_{i})~.
\ee
It is then a simple matter to substitute these functions into the general expression for the index of any $\cN=1$ theory. For illustration, we take the example of an unpunctured UV curve, where the index becomes
\be\label{CHLindexex}
\cI\left[\cC_{g}^{\ell_{1},\ell_{2}}\right]=\mbox{PE}\left(\sum_{j=2}^{k}\((j-1)(g-1)+j\(\frac{\ell_{2}-\ell_{1}}{2}\)\)\w^{j}\right).
\ee
The coefficient of $\w^{2}$ counts the relevant operators of the theory as discussed above. The number of relevant operators encoded in the index \eqref{CHLindexex} then takes the very elegant form
\be\label{allrelevant}
\#\ {\rm relevant}=(g-1)+|\ell_{2}-\ell_{1}|~.
\ee

The above result is tantalizing --- it matches a very natural geometric quantity relevant to the construction of these theories:
\be
\dim_\C H^1(\cC_g;{\rm ad}^{-}(P))=(g-1)+|\ell_2-\ell_1|~.
\ee
Here $P$ is the principal $SU(2)$ bundle whose connection describes the fibration of the normal bundle in the Calabi-Yau three-fold \eqref{CYgeom} over $\cC_g$, and ${\rm ad}^-(P)$ is the line bundle that forms the part of the adjoint bundle of $P$ with negative curvature. This is precisely the topological quantity which counts the \emph{Morse index} of a critical $SU(2)$ connection with respect to the Yang-Mills functional.\footnote{C.B. would like to thank A.J. Tolland for introducing and guiding him through the beautiful work of \cite{Atiyah:1982fa}, which was an experience that served to orient some of the present work.} 

The parallel between the structure of the class $\cS$ fixed points and the mathematics of two-dimensional $SU(2)$ Yang-Mills theory runs deeper than this, with the relation arising from the description of the local geometry of the Calabi-Yau in terms of an $SU(2)$ connection on the UV curve. In the holographic dual description of the fixed points \cite{Bah:2011vv,Bah:2012dg}, the critical values of the $SU(2)$ connection are precisely the Yang-Mills connections, \ie, those connections that are critical points of the Yang-Mills functional. Consequently, the conformal manifold of these theories was identified locally with the product of the complex structure moduli space of the curve with the critical manifold of the associated Yang-Mills connections (this is just the space of abelian Wilson lines in the general case). It is natural to expect that renormalization group flows among the fixed points at a given genus should be geometrized in the language of Yang-Mills connections and two dimensional geometry --- in particular, the holographic story should be an extension of \cite{Anderson:2011cz}, although since the spectrum of light operators is fairly insensitive to the rank of the theory, we may expect the picture to persist at finite $N$.

It is exciting to see hints of a famous application of Morse theory appearing in this context, which suggests a relationship between renormalization group flow in class $\cS$ and gradient flow of the Yang-Mills functional. From the holographic picture, it is clear that this will not be the whole story, since beyond leading order in conformal perturbation theory the metric on the UV curve will backreact, necessitating the inclusion of additional degrees of freedom in the flow. Nonetheless, the picture is appealing, and deserves further investigation.

\acknowledgments
It is a pleasure to thank Nikolay Bobev, Leonardo Rastelli, Shlomo Razamat, A.~J.~Tolland, and Wenbin Yan for numerous helpful discussions. The authors gratefully acknowledge the hospitality of the Aspen Center for Physics at the start of this work. The Aspen Center for Physics is partially supported by the NSF under Grant No. 1066293. The work of CB is supported in part by DOE grant DE-FG02-92ER-40697. The work of AG is supported in part by the John A. McCone fellowship and by DOE grant DE-FG02-92-ER40701.

\bibliographystyle{./files/JHEP}
\bibliography{./files/cjb_long}

\end{document}